\documentclass[11pt]{article}
\pdfoutput=1

\usepackage{amsmath, amsfonts, amssymb}
\usepackage{comment}
\usepackage{graphicx}
\usepackage{psfrag}
\usepackage{amsthm}
\usepackage[usenames,dvipsnames,svgnames,table]{xcolor}
\usepackage{enumerate}
\usepackage{tcolorbox}

\usepackage{soul}
\usepackage{subfig}
\usepackage{mathrsfs}  
 \usepackage{a4wide}
  \usepackage{tikz}
  \usepackage{tikz-cd}

  \usepackage{color}
  \definecolor{dark-gray}{gray}{0.20}
  \definecolor{gray}{gray}{0.30}
  \definecolor{light-gray}{gray}{0.80}
  \definecolor{dark-red}{rgb}{0.7,0,0}
  \definecolor{dark-green}{rgb}{0.1,0.4,0}
  \definecolor{dark-blue}{rgb}{0.3,0.3,0.7}
  \definecolor{light-blue}{rgb}{0.8,0.8,1}
      \definecolor{swamp}{RGB}{240, 199, 197}
  
  \usepackage{pifont}
\usepackage{setspace}

\newcommand{\be}{\begin{equation}}
\newcommand{\ee}{\end{equation}}
\newcommand{\eq}[1]{(\ref{#1})}

\def\be{\begin{equation}}
\def\ee{\end{equation}}
\def\bea{\begin{eqnarray}}
\def\eea{\end{eqnarray}}

\def\simleq{\; \raise0.3ex\hbox{$<$\kern-0.75em
      \raise-1.1ex\hbox{$\sim$}}\; }
   \def\simgeq{\; \raise0.3ex\hbox{$>$\kern-0.75em
      \raise-1.1ex\hbox{$\sim$}}\; }

\numberwithin{equation}{section}

\usepackage{jheppub} 
\hypersetup{
	colorlinks=true,
	linkcolor=dark-blue,
	citecolor=dark-red,
	urlcolor=dark-green,
	linktoc=page,
	pageanchor=false
}
\title{\centering The FL bound and its phenomenological implications}

\author{Miguel Montero$^1$,}
\author{Cumrun Vafa$^1$,}
\author{Thomas Van Riet$^{2,3}$}
\author{ and Gerben Venken$^4$}

\affiliation{$^1$Jefferson Physical Laboratory, Harvard University,\\
Cambridge, MA 02138, USA}
\affiliation{$^2$Institute of Theoretical Physics, KU Leuven,\\
Celestijnenlaan 200D B-3001 Leuven, Belgium
}
\affiliation{$^3$ Institutionen f\"{o}r Fysik och Astronomi,\\ Box 803, SE-751 08 Uppsala, Sweden} 
\affiliation{$^4$ Institute for Theoretical Physics, Heidelberg University,\\
Philosophenweg 19, 69120 Heidelberg, Germany} 
\emailAdd{mmontero@g.harvard.edu}
\emailAdd{vafa@g.harvard.edu}
\emailAdd{thomasvr@itf.fys.kuleuven.be}
\emailAdd{G.Venken@ThPhys.Uni-Heidelberg.DE}

\abstract{Demanding that charged Nariai black holes in (quasi-)de Sitter space decay without becoming super-extremal implies a lower bound on the masses of charged particles, known as the Festina Lente (FL) bound.
In this paper we fix the $\mathcal{O}(1)$ constant in the bound and elucidate various aspects of it, as well as extensions to $d>4$ and to situations with scalar potentials and dilatonic couplings.  We also discuss phenomenological implications of FL including an explanation of why the Higgs potential cannot have a local minimum at the origin, thus explaining why the weak force must be broken.  For constructions of meta-stable dS involving anti-brane uplift scenarios, even though the throat region is consistent with FL, the bound implies that we cannot have any light charged matter fields coming from any far away region in the compactified geometry, contrary to the fact that they are typically expected to arise in these scenarios.  This strongly suggests that introduction of warped anti-branes in the throat cannot be decoupled from the bulk dynamics as is commonly assumed.
Finally, we provide some evidence that in certain situations the FL bound can have implications even with gravity decoupled and illustrate this in the context of non-compact throats.}

  \preprint{UUITP-26/2}

\begin{document}

\makeatletter
\let\old@fpheader\@fpheader

\makeatother

\maketitle


\section{Introduction}\label{intro}

The Swampland program  \cite{Vafa:2005ui, Brennan:2017rbf, Palti:2019pca, vanBeest:2021lhn} aims to characterize the effective field theories that can arise as the low-energy limit of an Einsteinian theory of gravity.  The strength of the program comes from the myriad of string compactifications that support it, often in very nontrivial ways. Another appealing feature that is slowly being uncovered shows that Swampland constraints arise as consequences of general principles, such as absence of global symmetries in quantum gravity, holography, or considerations about black hole dynamics.   Moreover the various Swampland criteria seem to form an inter-connected web of ideas reinforcing one another and suggesting perhaps a unified set of principles.   The large amount of ``experimental'' evidence for Swampland constraints can be recast as evidence that these general principles are probably correct.  If so, we can apply them to situations which are currently beyond reach of controlled string compactifications, such as SUSY breaking situations like models of slow-rolling vacuum energy. This is how the Festina Lente (FL) bound for charged particles in dS space came about \cite{Montero:2019ekk}. The bound states that in dS space the mass for every state of charge 1 under a $U(1)$ gauge field with coupling $g$ satisfies 
 $$\frac{m^4}{8\pi \alpha}\geq V$$
where $m$ is the mass, $\alpha =\frac{g^2}{4\pi}$ is the fine structure constant and $V=\Lambda / 8\pi G$ is the gravitating vacuum energy density. The exact numerical coefficient of the bound was left undetermined in \cite{Montero:2019ekk}; in the present paper, we will see that consistency with the results of \cite{Huang:2006hc} fixes it to the value quoted above.

The  bound comes precisely from direct application of the principles behind the Weak Gravity Conjecture (WGC) \cite{ArkaniHamed:2006dz} to charged black holes in de Sitter space:  Reissner-Nordstr\"om-de Sitter black holes.
In the context of dS space the consistency of the decay of black holes much smaller than dS radius and avoiding super-extremality leads to the usual WGC which puts an upper bound on the mass of charged elementary states: $m<g M_{P}$.  Whereas if one considers large black holes whose size is comparable to dS radius and consider their decay, avoiding superextremality leads to the FL bound which is a lower bound on $m$.  In other words (assuming the WGC state is charge 1) we have both an upper and lower bound on its mass:
$$(8\pi \alpha\,  V)^{1/4}<m<{(8\pi \alpha)}^{1/2} M_{\text{P}}$$
This bound is nicely satisfied in our universe for the electron, giving a  weak quantitative prediction from Swampland principles. Note that aside from $\alpha$ the lower bound is close to the neutrino mass scale and the upper bound is the Planck mass. It is interesting to note that the lower and upper bounds on the mass together imply a lower bound on $\alpha$,
$$8\pi\alpha\gtrsim \frac{V}{M_{P}^4}.$$
Reassuringly this bound was previously proposed in \cite{Huang:2006hc} based on the magnetic form of the WGC in dS space.  This is a nice consistency check of the FL proposal.  In other words, the completeness of the magnetically charged spectrum (and the requirement that the corresponding BH fit in dS) leads to this bound.

The focus of \cite{Montero:2019ekk} was mainly on the general relativistic computation motivating the bound and only certain aspects of its phenomenological applications were touched upon. The aim of this paper is twofold.  On the one hand we would like to study the phenomenological implications in some more depth and on the other hand we wish to extend and generalize the FL bound to other dimensions as well as non-trivial rolling scalar potentials and dilatonic couplings. Along the way we find more motivation for the FL bound from stringy considerations.

Any Swampland bound that applies to light particles is relevant for phenomenology, and this is one of the main motivations behind our work. Since FL precludes the existence of light charged fields, there cannot be a phase of the Standard Model where the weak interaction is long range. As a consequence, there cannot be a local minimum at $\Phi=0$ for the Higgs potential. The other possibility consistent with non-abelian gauge fields and FL, confinement, is realized by the gluons. 

The scale set by the FL bound is tantalizingly close to the neutrino mass scale, but the bound does not apply since the neutrinos are uncharged. We discuss a simple microscopic scenario in which one can explicitly construct charged states satisfying the FL bound. In addition to these, in this scenario we find that these charged states are accompanied by neutral states at a similar mass scale.  It would be interesting to see if this holds more generally, potentially leading to an explanation of the neutrino mass scale.  For other connections between the Swampland principles and the neutrino mass scale see \cite{Ooguri:2016pdq,Ibanez:2017kvh,Hamada:2017yji,Ibanez:2017oqr,Lust:2017wrl,Gonzalo:2018tpb,Gonzalo:2018dxi,Gonzalo:2021fma,Rudelius:2021oaz}. 
 
This paper is organized as follows. In Section \ref{sec:rev} we review the FL proposal and discuss its generalizations to multi-field situations, towers of states, magnetic version, and discuss a plausible microscopic scenario where an FL-like bound appears automatically. In Section \ref{sec:dimred} we study the interplay of FL with dimensional reduction, and find that positive vacuum energy together with FL suggests that to stabilize a quasi-dS solution from one in higher dimension the stabilized radii should be smaller than Hubble scale. Section \ref{sec:ph} discusses phenomenological and model-building implications of the proposal, including neutrino physics and statements about the requirement of instability at top of the Higgs potential. Section \ref{sec:evi} discusses the non-compact limit of FL, where gravity is decoupled.
Section \ref{sec:app} studies FL and the WGC in anti-brane uplift scenarios, where the branes are localized in a deep throat of a compact geometry. We find that both statements are satisfied automatically, as a consequence of the decoupling of the throat dynamics. But FL also constrains sectors outside of the throat in the compact setup, which raises doubts about the basic logic underlying anti-brane uplift scenarios which ignores the coupling of the anti-brane with the bulk dynamics. Finally, we present our conclusions in Section \ref{conclus}. Some computations and additional results are relegated to Appendices. 

\section{The Festina Lente proposal}\label{sec:rev}

The main character in this paper is the Festina Lente (FL) conjecture of \cite{Montero:2019ekk}, so we start by reviewing the logic behind it, as well as its interplay with other Swampland constraints in de Sitter such as TCC \cite{Bedroya:2019snp} and the de Sitter  conjecture \cite{Ooguri:2018wrx}.

Consider a long-lived de Sitter solution, meaning a positive vacuum energy configuration in Einstein gravity such that there are no perturbative instabilities, and with a lifetime of order the Hubble scale or longer (we do not expect it can be longer than $(1/H) \log H$, as this would violate the TCC \cite{Bedroya:2019snp}). We are interested in the case where the low-energy EFT contains a long-range $U(1)$ force. As we will explain below, there is no loss of generality in considering the abelian case.

For simplicity, we start discussing a single $U(1)$ gauge field with no dilaton coupling. Extensions to the case of multiple $U(1)$'s and scalar fields are discussed further below.  The low-energy EFT for the gauge-gravity sector is just the $d$-dimensional Einstein-Maxwell Lagrangian,
\begin{equation}S=\int d^dx\sqrt{-g}\left[ \frac{1}{2}M_p^{d-2} \mathcal{R} - \frac{1}{4g^2}F_{\mu\nu}F^{\mu\nu} - V\right].\label{s0}\end{equation}
Here, $M_p^2=\frac{8\pi}{G}$ and $V$ is the  cosmological constant/vacuum energy, related to the de Sitter radius  $\ell\equiv 1/H$ as
\begin{equation} \frac{(d-1)(d-2)}{2\ell_d^2}=M_p^{2-d}\, V.\end{equation}
Such an EFT, if valid, assumes full moduli stabilisation. 

It is by now a familiar fact that consistency with quantum gravity imposes nontrivial constraints on the spectrum of charged states in the theory. In particular, we must impose the Weak Gravity Conjecture (WGC), proposed in \cite{ArkaniHamed:2006dz} and further studied in e.g. \cite{Heidenreich:2016aqi,Montero:2016tif,Hamada:2018dde,Cheung:2014vva,Palti:2017elp}, which bounds the charge-to-mass ratio of charged particles, 
\begin{equation} \frac{gq}{m}\geq \sqrt{\frac{d-3}{d-2}}M_p^{-\frac{(d-2)}{2}}\quad  \text{for some charged state in the theory}.\end{equation}
Here, $q$ is the integer-quantized electric charge of the particle. This bound is related to the evaporation of small, near-extremal charged black holes \cite{ArkaniHamed:2006dz}. On the other hand, the arguments in \cite{Montero:2019ekk} suggest that, for positive vacuum energy, we must impose the additional condition
\begin{equation} m^4\gtrsim (gq)^2V\quad \text{for every charged state in the theory}.\label{FL00}\end{equation}
Equation \eq{FL00} is the FL proposal. Note that the inequality \eq{FL00} remains the same in any number of dimensions, since the gauge coupling has units of $\text{Energy}^{2-d/2}$. Also, in the original analysis of \cite{Montero:2019ekk}, the $\mathcal{O}(1)$ factor is left undetermined, since it comes from the details of the spectrum and the Schwinger effect at strong coupling. In Section \ref{subsec:mch} of the present paper, we will see how it can be fixed by consistency with the analysis of \cite{Huang:2006hc}.

\begin{figure}[!htb]
	\begin{center}
		\includegraphics[width=0.6\textwidth]{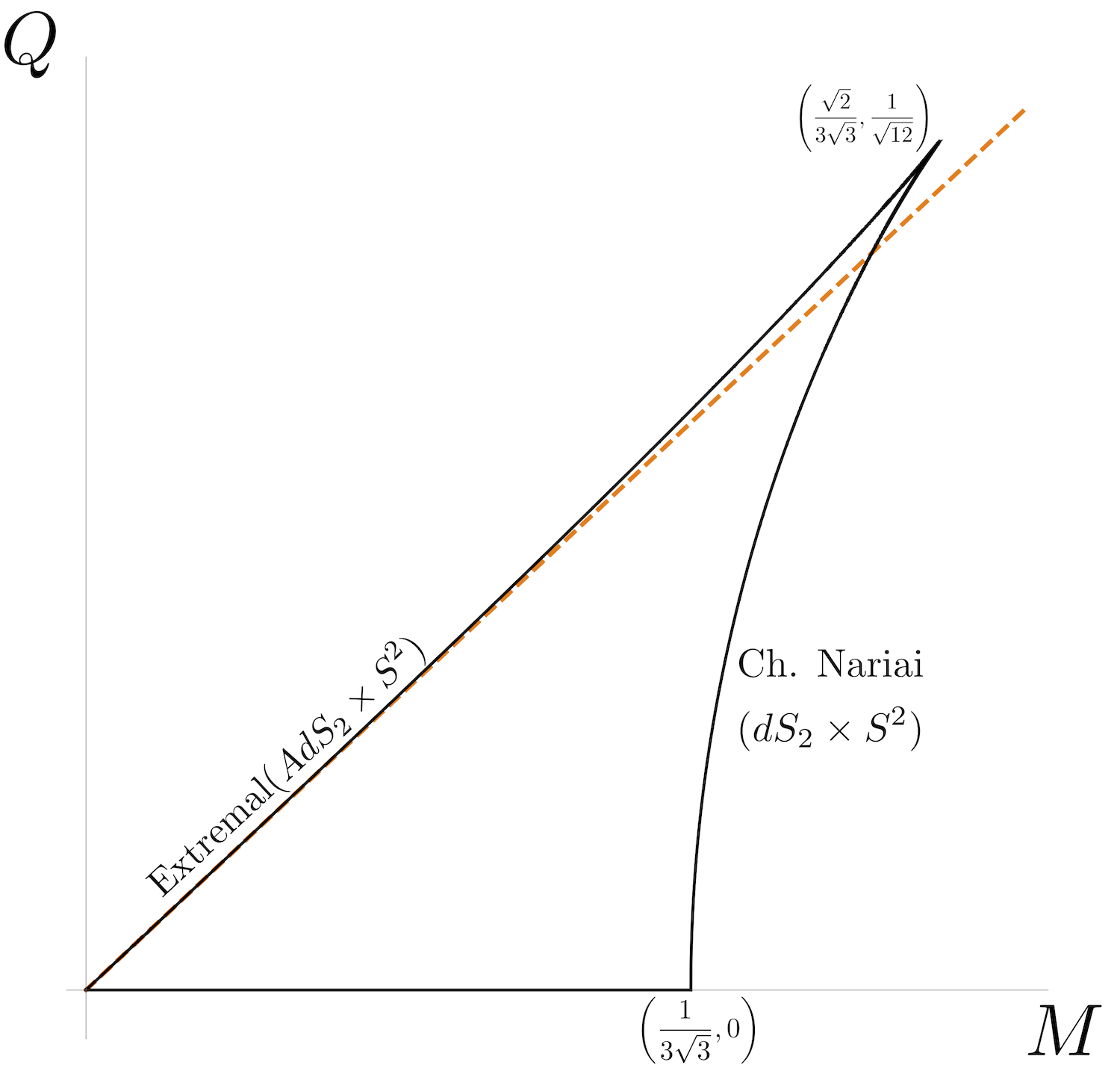}
		
		\caption{The family of Reissner-Nordstr\"om-de Sitter black holes with the dimensionless parameters $M\equiv \frac{GM}{\ell_4},\quad Q^2\equiv \frac{Gg^2Q^2}{4\pi \ell_4^2}$ with $\ell_4$ the dS length $H^{-1}$. Sub-extremal solutions only exist inside the shark-fin shaped region. The boundary of this allowed region has two branches: On the left branch one has extremal RN-dS black holes, which have a $AdS_2\times S^2$ horizon. On the  right branch one has charged Nariai black holes for which the black hole and cosmological horizons coincide and become a $dS_2\times S^2$. The orange dashed line is the ``lukewarm line'' $Q=M$, where the two horizons have the same temperature. Picture taken from \cite{Montero:2020rpl}.}\label{fig1}
	\end{center}
\end{figure}

Just like the WGC, \eq{FL00}  is supported by arguments coming from black hole physics, which we now review. For simplicity of the presentation we now carry on with $d=4$. 
The action \eq{s0} admits charged black hole solutions known as the Reissner-Nordstr\"om de Sitter family:
\begin{align}
ds^2 & = -U(r)dt^2 + \frac{dr^2}{U(r)} + r^2d\Omega_2^2\,,\\
F_2 &  = \frac{g^2}{4\pi}\frac{Q}{r^2}dt\wedge dr\,, 
\end{align}
where
\be
U(r)= 1 - \frac{2GM}{r} + \frac{G(Qg)^2}{4\pi r^2} -\frac{r^2}{\ell_4^2}\,.
\ee 
In contrast to Minkowski and AdS black holes, which can be arbitrarily large, there is a maximum size that a black hole in de Sitter space can have.  Physically, this limitation comes about because a static black hole solution must fit within its own cosmological horizon; this means that the black hole horizon is smaller than the cosmological one. See Figure \ref{fig1}.

 The limit where the two approach each other is realized by a near-horizon geometry of the form $dS^2\times S^2$, also known as the (charged) Nariai metric \cite{Nariai1999OnSS,Romans:1991nq}
\begin{align} ds^2 &=- \left(1 -\frac{\rho^2}{\ell_2^2}\right)d\tau^2+ \left(1 -\frac{\rho^2}{\ell_2^2}\right)^{-1} d\rho^2+ r_c^2d\Omega^2\,,\label{metric-nariai} \\
F_2 & =   \frac{g^2}{4\pi} \frac{Q}{r_c^2} d\rho \wedge d\tau\,,
\end{align} 
where 
\begin{align}
\ell_2^2 = \left(3-\frac{Gg^2Q^2}{4\pi \ell_4^2r_c^4}\right)^{-1}\ell_4^2  \,,\qquad r_c = \frac16\left(1+\sqrt{1-12 \frac{Gg^2Q^2}{4\pi \ell_4^2}}\right)\ell_4\,. 
\end{align}
These charged Nariai solutions are a family of classically stable, ``extremal'' black hole solutions \cite{Romans:1991nq}.

The WGC implies that nonsupersymmetric charged black holes must be kinematically allowed to evaporate \cite{ArkaniHamed:2006dz} and FL comes from applying a similar principle to the Nariai solutions. Including charged matter causes evaporation of the charged Nariai solutions as analyzed in detail in \cite{Montero:2019ekk, Luben:2020wix}. The decay is triggered by Schwinger pair production in the near-horizon electric field, which is controlled by a rate
\begin{equation} \Gamma \sim \exp\left(-\frac{m^2}{qE}\right)= \exp\left(-\frac{m^2}{(g q \sqrt{V})}\right).\label{sch3}\end{equation}
In the last inequality, we have substituted the typical electric field of the Nariai black hole, which is of order $g M_P H= g \sqrt{V}$. As shown in \cite{Montero:2019ekk}, the fate of the black hole is very different depending on whether \eq{sch3} is exponentially suppressed or not. There, it was found that if the charged matter is very light, such that  \eq{FL00} is not satisfied, charged Nariai black holes do not evaporate back to empty de Sitter space; instead, they crunch and develop arbitrarily high curvatures in a time of order $1/H$. This is because their electric field is quickly screened by Schwinger pair production. By contrast, if  \eq{FL00} is satisfied, then all black holes slowly evaporate towards empty de Sitter space in the usual fashion. Demanding that the first case does not take place then leads to \eq{FL00}.

We do not usually throw away a theory just because a particular solution happens to crunch or otherwise become singular. However, as studied in \cite{Montero:2019ekk}, charged Nariai black holes crunching in this way are effectively becoming super-extremal, having more mass than that of the neutral Nariai black hole -- more mass than what can be fit in the de Sitter static patch. For near-extremal black holes in Minkowski, avoidance of naked singularities and superextremal solutions leads to the WGC \cite{ArkaniHamed:2006dz}. Avoiding a similar pathology in dS suggests that one should impose \eq{FL00} as a Swampland constraint. Thus the FL bound, together with the weak gravity bound, ensure that black holes never become super-extremal in de Sitter space.

Nariai black holes are an essential ingredient in the derivation of the bound. They always exist as solution to the equations of motion, and are continuously connected to smaller black holes, so there is no obvious way to get rid of them without introducing an arbitrary cut-off in charge or configuration space. On top of this, they can be produced from the vacuum via nonperturbative effects. In particular, see the discussion around page 15 of \cite{Bousso:1999ms} and references therein, which describe a family of instanton solutions in the pure Einstein-Maxwell theory that interpolate between empty $dS$ space and charged Nariai solutions. It would be interesting to study the fate of these instantons when light charged matter is present. 

Finally, we would like to offer a speculation. The WGC is connected to Weak Cosmic Censorship (WCC) \cite{Crisford:2017gsb,Horowitz:2019eum}, the idea that there should be no naked singularities that are not cloaked by horizons. These would appear if an extremal black hole could evaporate by emission of sub-extremal particles, becoming superextremal. Similarly, the Big Crunch singularities in an overextremal Nariai solution are visible to every observer, and not cloaked by a horizon (unlike the curvature singularity of a Schwarzschild black hole). So if an extension of WCC ideas forbids spacelike singularities outside horizons, that would provide a rationale for the FL bound (and much more, as it would restrict the allowed set of initial conditions in GR).

\subsection{FL bound for runaway potentials}

As explained above, in order to apply the FL bound we require the existence of a charged Nariai black hole solution. Because of this \cite{Montero:2019ekk}, FL focused on the case of a quasi-de Sitter background, with no perturbative instabilities and a lifetime of a Hubble time or larger, such as our own universe\footnote{In such a scenario, the bad crunching singularity of the superextremal Nariai geometry is reached before the instability of the parent dS vacuum can be triggered.}. However, even a fast runaway scalar potential can still admit Nariai solutions under certain conditions, as pointed out in \cite{Montero:2020rpl} and we recall now. This allows for a generalization of FL to certain cases with runaway potential.

Consider a four-dimensional theory with a scalar rolling in a potential $V>0$ (an inflaton or quintessence field) and a $U(1)$ gauge field. Generically, the scalar $\phi$ will couple to the gauge field and we obtain an action of the form\footnote{One may add to this action a term $\frac{\theta}{8\pi^2}  F\wedge F$. In this case the bounds that follow also depend on $\theta$, see \cite{Montero:2020rpl} for further details. One can then rewrite the equations in terms of $\tau=2\pi i f(\phi) +\frac{\theta}{2\pi}$.

}
 \be \label{quintessenceaction}
S= \int\sqrt{|g|}\Bigl(\frac{1}{2}M_p^2 \mathcal{R}-\tfrac{1}{2}(\partial\phi)^2 -\tfrac{1}{4}f(\phi) F_{\mu\nu}F^{\mu\nu} - V(\phi)\Bigr) + \text{matter} \,.
\ee
The construction, or even existence, of black hole solutions within such a theory can be involved, see for instance \cite{Benakli:2021fvv}. In dS space the boundary of the diagram of black hole solutions in dS space consists of (AdS$_2$, Mink$_2$, $dS_2)\times S^2$ solutions with the Minkowski solution forming the crossover from $AdS_2$ to $dS_2$.  Similarly here, the boundary can be explicitly constructed simply from an $S^2$ compactification with electric and magnetic fluxes (charges) as shown in \cite{Montero:2020rpl}. The Nariai solutions would then correspond to the $dS_2\times S^2$ branch. If one is to choose freely both magnetic and electric black hole charges than there will be always a Nariai branch, as long as the following inequality is fulfilled:
\begin{equation}\left\vert\frac{V'}{V}\right\vert < \left\vert\frac{f'}{f}\right\vert. \label{dScond0}\end{equation}
Note that this is a necessary requirement for application of FL and seems to be in tension with $dS$ conjecture \cite{Ooguri:2018wrx}. However, it could be that this is still compatible with $dS$ conjecture as long as the right hand side of the above inequality is bigger than the $O(1)$ number in the statement of the $dS$ conjecture.\footnote{Note that most of the evidence for $dS$ conjecture comes from weak asymptotic regions of parameter space and it is conceivable that the $dS$ conjecture is replaced by TCC \cite{Bedroya:2019snp}  more generally which would be compatible with the above inequality.}

Even if we have a $dS_2 \times S^2$ solution, we require it to be classically stable under scalar perturbations for our arguments against the existence of light charged matter to apply. In case the Nariai black hole is purely electric (when $f'$ and $V'$ have the same sign) perturbative stability requires \cite{Montero:2020rpl}
\begin{equation}V'' > \frac{V'}{f'}\left( f''-2\frac{f'^2}{f}\right) .\label{stab}\end{equation}

The generalization of these results to higher dimensions is straightforward, and is carried out in Appendix \ref{App:higherd}. Here we merely quote the results: A $d$-dimensional theory has Nariai-like solutions of the form $dS_2\times S^{d-2}$ supported by electric flux through the $dS^2$  only if the potential and gauge kinetic functions satisfy
\begin{equation}\frac{V'}{V}< (d-3) \frac{f'}{f} \qquad \text{and}\qquad \text{sign}(V')=\text{sign}(f')\label{ineqd}\,.\end{equation}
Equation \eq{stab} on the other hand remains unchanged.

In particular, notice that \eq{ineqd} cannot be satisfied for $d\leq 3$. As it is often the case whenever gauge or gravitational fields are involved, the three-dimensional case is special and set apart from the others; there are no Nariai solutions.  For $d\geq 4$, if a Nariai solution exists because (\ref{ineqd}) is satisfied and if it is perturbatively stable because (\ref{stab}) is satisfied, we can use the same argument against the existence of light charged matter arriving at the FL bound (\ref{FL00}). We therefore arrive at the full statement of the FL bound that we use in this paper, and box it for easy referencing\footnote{Although it does not follow from the analysis we have presented so far, we have also substituted the precise $\mathcal{O}(1)$ coefficient that follows from consistency with magnetic versions of the WGC in de Sitter, which will be analyzed in Section \ref{subsec:mch}. }:
\vspace{2mm}

\begin{tcolorbox}
\textbf{The Festina Lente (FL) bound}: If the signs of $V'$ and $f'$ are the same, and the inequalities
\begin{equation} \frac{V'}{V} \leq (d-3) \frac{f'}{f} \quad\text{and}\quad 
V'' \geq \frac{V'}{f'}\left( f''-2\frac{f'^2}{f}\right) \end{equation}
are satisfied, there exist classically stable electric Nariai solutions to which we can apply the bound that  \textbf{every} particle of charge $q$ and mass $m$ must satisfy the inequality
\begin{equation} m^4\gtrsim 6 (g q M_P H)^2 = 2(gq)^2V\label{FL0}.\end{equation}
\end{tcolorbox}
\vspace{2mm}

We finish with a technical note. For general $d$, the existence condition (\ref{dScond0}) comes from analyzing an electric Nariai solution, with topology $dS_2\times S^{d-2}$. For this solution to exist, it is necessary that the signs of $f'$ and $V'$ are the same. It is often the case in asymptotic limits in string compactifications that there is a limit where both the vacuum energy and the gauge coupling vanish asymptotically \cite{Gendler:2020dfp, vanBeest:2021lhn}. In the special case of four dimensions, however, one can make the argument with both the electric and magnetic $U(1)$'s. In this case, by considering solutions with magnetic charge, we can ensure there is always a Nariai solution, irrespectively of the relative sign of $V'$ and $f'$ (if \eqref{dScond0} is obeyed). If the black hole has electric charge as well, and is comparable to the magnetic one, the evaporation analysis of \cite{Montero:2019ekk} still holds. Furthermore, in the case where only magnetic fields are involved, a more heuristic argument involving instability via pair production of magnetic dipoles can be made, with the same parametric behavior as \eq{FL0} (see Section \ref{sec:ph}). To sum up, only in $d=4$, the condition that $f'$ and $V'$ have the same signs in \eq{FL0} can be dropped, if we allow for consideration of dyonic black holes. 

\subsection{Generalisations of the FL bound}\label{secgeneralisation}
As we discussed in the introduction, one of the main points of this paper is to give the FL inequality the same treatment as any other Swampland constraint -- and work out its generalizations in various situations, such as multiple fields, or the magnetic version.  We discuss a few of these in the following:

\subsubsection{FL-bound for a tower of charged particles }
\label{secFLtowerbound}

UV completions of effective field theories sometimes imply that a state in an EFT is the lightest state of an infinite tower \cite{Palti:2019pca, Brennan:2017rbf, vanBeest:2021lhn}. Hence it makes sense to check whether the FL bound is changed in the presence of a tower of charged states. Intuitively, a tower amplifies the Schwinger-pair process since more particles are taking part in the decay process.  For a single particle species,  the Schwinger pair production rate is governed by the exponential \eq{sch3}. But more generally, we should take into account every particle species, and impose $\Gamma\ll 1$.
In presence of a tower of unit charged particles with degeneracies $N(m_i)$ per mass $m_i$ this becomes
\be \label{towerFL}
\Gamma \sim \sum_i N(m_i)e^{-\frac{m_i^2}{qgHM_p}}\quad\rightarrow \quad\int dm \rho(m)e^{-\frac{m^2}{qgHM_p}}\ll1,
\ee
where the last step introduces a density function in the continuum limit. Later in this paper we will discuss string theoretic examples of the FL bound from dS constructions based on warped throats and then this tower condition will be required since warping lowers the mass of otherwise heavy particles within a tower.

\subsubsection{Multi-field generalization}

The FL bound can also be straightforwardly generalized to the case of several $U(1)$ fields. This multi-field generalization is discussed in more detail in Appendix \ref{App:multi} and reads :\begin{equation}
m^4 >> q_Aq_B (f^{-1})^{AB} (M_pH)^2 \,,
\end{equation}
where $f_{AB}$ appears in the gauge kinetic function, $-\tfrac{1}{4} f_{AB}(\phi)F^AF^B$, and $q_A$ is the electric charge vector of the particle with mass $m$. Further extensions to multi-field theories with dyonic particles can also be found in the Appendix \ref{App:multi}. 

\subsubsection{Magnetic version of the FL inequality}\label{subsec:mch}
The WGC comes in two versions: an electric version constraining the mass spectrum of the theory, and a magnetic version constraining the cut-off scale of the EFT. Since the FL inequality (\ref{FL0}) constrains the masses of the particles in the theory, it resembles an ``electric'' version. We now wish to formulate its magnetic version. 

The magnetic version of the WGC \cite{ArkaniHamed:2006dz} states that the UV cut-off scale of an EFT is bounded by
\begin{equation}
\label{eqMagWGC}
\Lambda_{EFT} \leq g M_{P}\,.\end{equation}

This was originally argued by demanding that the monopole state of unit charge is outside its own horizon, i.e. that it is not a black hole. In \cite{Huang:2006hc}, this rationale for the magnetic WGC was extended to four-dimensional de Sitter space. By demanding that a monopole of charge one should never be a (Nariai) black hole, \eq{eqMagWGC} is modified in such a way that a precise lower bound in the gauge coupling can be obtained. We now review the argument\footnote{The analysis in this section is different from that in \cite{Huang:2006hc}, since they used the neutral Nariai black hole instead of the magnetically charged one to estimate the radius. This produces the same parametric dependence, so the qualitative conclusions are the same, but getting the factors right allows one to get a precise bound on the gauge coupling, which we will use to fix the $\mathcal{O}(1)$ factors in FL. }. Consider a magnetic monopole of mass $M$ and radius $1/\Lambda_{EFT}$, given by the inverse cut-off of the effective field theory. In \cite{ArkaniHamed:2006dz} and \cite{Huang:2006hc} the estimate $M\sim\Lambda_{EFT}/g^2$ is used, which introduces uncontrolled $\mathcal{O}(1)$ factors and neglects contributions coming from the core of the monopole, which could be larger, or even negative, as the tension of orientifold planes in string theory. In de Sitter space, as we will see, we can do better.

The basic argument behind the magnetic WGC in \cite{ArkaniHamed:2006dz} is that the monopole of charge one should be a fundamental particle, i.e. it should not be a black hole. As a result, the exterior field of the monopole (beyond $r\sim 1/\Lambda_{EFT}$) should be described by a subextremal magnetically charged Reissner-Nordstrom-dS solution (see  Section \ref{sec:rev}). Following \cite{Huang:2006hc}, the polynomial
\begin{equation} P(r,\tilde{M},\tilde{Q})\equiv-r^2 U(r)= \frac{r^4}{\ell_4^2} + 2\tilde{M}r -\tilde{Q}^2n^2-r^2 \end{equation}
 must be $\leq0$ when evaluated at $r=\Lambda_{EFT}^{-1}$. We have defined
\begin{equation}\tilde{M}= GM,\quad \tilde{Q}^2 n^2= \frac{G}{4\pi} \frac{Q_m^2}{g^2}=  \frac{G\pi }{g^2} n^2\end{equation}
in terms of the integer monopole charge $n$. Setting $n=1$,  the extension of the magnetic WGC argument to RN-dS black holes yields an implicit equation
\begin{equation} P(\Lambda_{EFT}^{-1},\tilde{M},\tilde{Q})\leq0.\end{equation}
For a given $\tilde{M},\tilde{Q}$, the above equation gives $\Lambda_{EFT}^{-1}$ implicitly, as the horizon radius of the corresponding black hole in Figure \ref{fig1}. This horizon radius is certainly larger than that of the extremal black hole, and smaller than that of the Nariai black hole. So in de Sitter we get both lower and upper bounds on the cut-off,
\begin{equation}\label{magWGCfordSBHbound} 
\frac{r_+}{\ell_4}=\sqrt{\frac16\left(1-\sqrt{1-12\tilde{Q}^2/\ell_4^2}\right)}\leq \frac{1}{\Lambda_{EFT} \ell_4} \leq\sqrt{\frac16\left(1+\sqrt{1-12\tilde{Q}^2/\ell_4^2}\right)}=\frac{r^{\text{Nariai}}_c}{\ell_4}.
\end{equation}
The lower bound becomes the usual magnetic WGC, if $\tilde{Q}$ is small enough. On the other hand, for $\tilde{Q}$ large, the two bounds become comparable, and the magnetic WGC cut-off is much smaller than its flat space counterpart. There is a sharp value, defined by
\begin{equation}\frac{\tilde{Q}^2}{\ell_4^2}=\frac{1}{12},\label{qmax2}\end{equation}
beyond which no magnetic Nariai solutions exist. This means that the coupling is so weak that not even the fundamental magnetic monopole fits in the static patch. Thus, the magnetic WGC argument in dS produces a sharp bound on the gauge coupling, obtained by rearranging \eq{qmax2}, which is simply
\begin{equation} g^2\geq \frac{3}{2} \left(\frac{H}{M_P}\right)^2.\label{HuangOK}\end{equation}
Parametrically, this is the same as the bound in \cite{Huang:2006hc}, and it is also equivalent to demanding that the ordinary flat-space magnetic WGC cut-off $g M_P$ should not be below the Hubble scale\footnote{We thank Irene Valenzuela for pointing this out to us.}.

In the case of the WGC, the magnetic WGC scale argued as above also coincides with the mass of the WGC particle, showing that the two are sides of the same coin. Interestingly, the same happens with FL and \eq{HuangOK}, modulo $\mathcal{O}(1)$ coefficients. FL has to be satisfied for every particle in the spectrum, and in particular it must be satisfied for a WGC particle of unit charge, which has $m\lesssim g M_P$. Combining these two, one gets
\begin{equation} m\lesssim gM_P\,\quad \text{and}\,\quad m^2\gtrsim g M_PH \quad\Rightarrow\quad g\gtrsim\frac{H}{M_P},\label{gr2}\end{equation}
so we also recover \eq{HuangOK}. This provides a mild consistency check of the FL proposal, since there is the same relationship between electric/magnetic arguments as there was for electric/magnetic WGC. But more interestingly, it offers an opportunity to fix the undetermined $\mathcal{O}(1)$ factor in the FL bound. Requiring that the derivation \eq{gr2} reproduces \eq{HuangOK}, together with a sharp formulation for electric WGC obtained from the exact extremality condition of small electric RN black holes, which is $M\leq\sqrt{2} g M_P Q$ (and we remind the reader that in our conventions $M_P^2=(8\pi G)^{-1}$), fixes the FL bound to 
\begin{equation} m^2 \geq \sqrt{6}\, g M_P H.\end{equation}
It would be very interesting to test this concrete prediction for the $\mathcal{O}(1)$ coefficient against the result coming from a detailed analysis of the particle production process via the Schwinger effect in the charged Nariai geometry. We also notice that $\sqrt{6} g M_P H$ coincides with the electric field of the ultracold Nariai black hole, the one at the tip of the shark fin in Figure \ref{fig1}, where the extremal and Nariai branches meet. The physics of this point, which has a local $\text{Mink}_2\times S^2$ near-horizon geometry, should be studied more thoroughly. 

Expanding the left hand bound in \eq{magWGCfordSBHbound} in $\tilde{Q}^2/\ell_4^2$ one finds
\begin{equation}
\Lambda_{EFT} < \sqrt{8} g M_P - \frac{3}{2} \frac{H^2}{\sqrt{8} g M_P} - \mathcal{O}\left(\frac{H^4}{M_P^3}\right)\,.
\end{equation}
The leading term reproduces the flat space magnetic WGC \eq{eqMagWGC}. We can now interpret the FL bound \eq{FL0} as an upper bound on $g$ and use this to obtain
\begin{equation} \label{magchargedparticlebound}
\Lambda_{EFT} < \frac{2}{\sqrt{3}}\frac{m^2}{H} - \frac{3 \sqrt{3}}{4} \frac{H^3}{m^2} - \mathcal{O}\left(\frac{H^7}{m^6}\right)\,.
\end{equation}
We see that the EFT cut-off scale is bounded by the mass of the charge carriers in the theory\footnote{This is reminiscent of the proposed bound $\Lambda_{EFT} \leq \sqrt{g m M_{P}}$ for pure QED in  \cite{Alberte:2020bdz,Aoki:2021ckh}. In this case too the charge carriers provide a (much stronger) upper bound on the EFT cut-off scale, albeit stemming from non-trivial assumptions about the scattering amplitudes. Note that the cut-off in \cite{Aoki:2021ckh} is dependent on the complete particle spectrum of the theory and is for instance heavily modified going from QED to electroweak theory.}. If we fill in real world numbers for the electron $m^2_e \sim 10^{11} \text{eV}^2$ and $H\sim10^{-33} \text{eV}$, we find a bound $\Lambda_{EFT} \leq 10^{35}\, \text{GeV}$, far above the Planck scale. Even employing the neutrino mass, the bound is a few orders of magnitude above Planck, so the bound \eq{Gbound} lacks interest unless one has very light charged states in the theory. The bound \eq{magchargedparticlebound} is independent of $g$. Thus, we can entertain the idea of applying it even to spontaneously broken symmetries, although we do not have a solid justification for this. For instance, for $\Lambda_{EFT}\sim M_{\text{GUT}}\sim 10^{15}\, \text{GeV}$, a particle saturating the bound would have  a mass around $10^{-5},\ eV$, a couple orders of magnitude below the neutrino mass scale. 
Requiring the left hand bound in \eq{magWGCfordSBHbound} to be real and using \eq{FL0} as an upper bound on $g$, we obtain 
\begin{equation}
\label{Gbound}
m^2>3\, H^2\,=\frac{{V}}{M_P^2},
\end{equation}
the mass of the lightest charged particles must remain heavier than the Hubble scale for the EFT to be sensible, a statement which is independent of $g$.

\subsubsection{Rotating black holes}
Nariai black holes that rotate instead of being charged are a good testing ground for the general principle behind the Festina Lente idea. Morally speaking, a rotating Nariai black hole is very similar to a charged electric one, except that electric charge is replaced by angular momentum. The loss of electric charge via pair production is replaced by a loss of angular momentum via emission of radiation, a process that can happen either classically (superradiance) or quantum-mechanically. One might then worry that e.g. maybe the orbital modes of the graviton, which have effective two-dimensional masses controlled by $H$, might be enough to trigger quick discharge of the black hole. However, this is not what happens; classically, superradiance shuts off in the Nariai limit \cite{Anninos:2010gh,Anninos:2012qw}. Quantum-mechanically, there can still be pair-production of particles due to the time-dependent metric, and these particles can take away angular momentum. The proper way to analyze the problem is to decompose four-dimensional fields in spherical harmonics, and study particle production in the effective two-dimensional geometry. Reference \cite{Anninos:2010gh} did this, and found that the equations of motion are very much the same as those for an electrically charged particle in $dS_2$, with an effective 2d electric field $E\propto H^2$
that scales like $H^2$ instead of $H$ as in the charged case. This means that, by following the same logic as in \cite{Montero:2019ekk}, one can impose \eq{FL00}, but $M_pH$ gets replaced by $H^2$. As a result, the bound becomes
\begin{equation} m_{\text{2d}}\gtrsim H, \label{h03}\end{equation}
and all 2d modes whose effective masses are $\gtrsim H$ satisfy the rotating version of the FL bound (although a mass of order Hubble means the corresponding field is close to saturating the bound; it would be interesting to explore how sharply the bound needs to be obeyed for consistency). A mass of order $H$ is also the lowest energy that can be measured in de Sitter space; intuitively, to measure energies below $H$, one would need to set up an experiment bigger than the cosmological horizon. Relatedly, modes whose effective mass is below $H$ are frozen by Hubble friction; and empty $dS$ space has a thermal bath at temperature $H/2\pi$ \cite{Birrell:1982ix}. Thus, on physical grounds, a bound like \eq{h03} is satisfied automatically in de Sitter space. 

Equation \eq{h03} in particular applies to the orbital modes of the graviton or a $U(1)$ gauge field, which will not trigger a catastrophic decay. Thus, the FL bound behaves very much like the WGC, in the sense that there is no version of it involving angular momentum instead of $U(1)$ charge,  because Kerr black holes can already decay towards sub-extremality by emitting gravitational radiation. 

The above arguments are qualitative. But \cite{Gregory:2021ozs} recently computed numerically the spindown of an extremal Kerr black hole, taking into account quantum effects due to gravity, scalar, and electromagnetic sources, finding that the black holes evaporate smoothly towards empty de Sitter space, similarly to the quasistatic discharge of electric black holes in \cite{Montero:2019ekk}. These results confirm the arguments in this Section.

\subsection{The FL bound from domain wall dynamics}\label{subsec:mem}
Above we described the ``macroscopic'' motivation for FL, coming from the decay of Nariai black holes, as outlined in \cite{Montero:2019ekk}, making the similarities to the WGC apparent. Unlike for the WGC, however, we do not understand the microscopic reasons for FL even in a single example, because we do not have a controlled dS solution in string theory. There is probably not a single, universal microscopic mechanism, just like for the WGC, which is realized due to different microscopic reasons in each particular model. For instance, in heterotic compactifications, the WGC can be traced to the $-1$ in the left-mover zero-point energy of the heterotic worldsheet, or due to the size of the cycles wrapped by branes, for RR $U(1)$'s \cite{ArkaniHamed:2006dz}.

In this Subsection, we entertain a particular class of models where a microscopic derivation of the physics behind FL can be analyzed.  While we do not expect it to be completely general, it will cover some of the stringy evidence that we discuss in Section \ref{sec:app} in particular for anti-brane uplifting. 

The scenario that we consider is that of a hypothetical quasi-de Sitter, supported by a top-form flux. This means that the effective four-dimensional action contains a term
\begin{equation}S\supset -\frac12\int \vert F_4\vert^2,\end{equation}
similarly to the Bousso-Polchinski scenario \cite{Bousso:2000xa}. We can also allow for several top-forms, but will restrict to one for simplicity. The 4-form $F_4$ takes on a quantized vev, $F_4=g_3 n$, and the vacuum energy is simply
\begin{equation} V=\frac12 g_3^2n^2,\quad n\in\mathbb{Z}.\end{equation}
In such a scenario, completeness of the spectrum \cite{Horowitz:1996nw} requires the presence of membranes that mediate transitions 
\begin{equation}n\rightarrow n'\end{equation}
and which allow the top-form flux to discharge, via the Brown-Teitelboim nucleation process \cite{Brown:1988kg}, hence making the de Sitter space decay (see \cite{Bedroya:2020rac} for an effective potential description of the case where the transitions happen very quickly). When the critical radius of the bubble is much smaller than the size of de Sitter, we can reliably use the flat-space formula \cite{Coleman:1980aw}
\begin{equation} \Gamma\sim \exp\left(-\frac{27\pi^2}{2}\frac{T^4}{(\Delta V)^3}\right),\label{cdl}\end{equation}
where $\Delta V$ is the change in vacuum energy and $T$ the tension of the bubble wall. This is the four-dimensional result; in $d$ dimensions, the action goes as $T^d/(\Delta V)^{d-1}$.

 It is often the case in string theory that the membranes that mediate vacuum decay have worldvolume fields of their own; for instance, if $F_4$ is a RR flux, the domain wall will be a D-brane, which will contain worldvolume gauge fields, and often these worldvolume gauge fields can turn on additional spacetime charges on the membrane, due to the existence of topological couplings in their worldvolume.
 This is the case we will be interested:  Consider a membrane with a worldvolume $U(1)$ gauge field $\tilde{A}$ and a coupling 
\begin{equation} \int_{\text{Membrane}} A \wedge \tilde{F}\,,\label{mcop}\end{equation}
in its worldvolume, where $A$ is the bulk $U(1)$ photon we are considering. By turning on $\tilde{F}$, a spherical membrane (wrapped on a contractible $S^2$ in a spatial slice of de Sitter) gets electric charge under the $U(1)$, thus producing a kind of ``membrane-particle''. This can also be understood as the statement that the electrically charged particle can polarize into a membrane via the Myers effect \cite{Myers:1999ps}.\footnote{One can wonder whether this argument can be repeated in other dimensions since the right kind of couplings might not exist. As explained in \cite{Gautason:2015tla} NS5 (and KK5) branes in IIA/IIB tend to have various forms of different ranks on their worldvolume that can be given a flux to provide the right kind of couplings for brane polarisation effect to occur.}

We will now estimate the mass of this ``membrane-particle'' that can be obtained by considering a spherical membrane with one unit of worldvolume flux included. Again, we emphasize that in stringy setups, this is nothing but a polarized brane with finite size due to the Myers effect. The charge prevents the membrane from collapsing, via its electrostatic repulsion, and classically stabilizes it at a radius (in four dimensions)
\begin{equation}
R \sim  \left(g^2/2T\right)^{1/3}\,.\label{r4}
\end{equation}
For small coupling, this radius is below the typical Compton wavelength of a membrane state, $T^{-1/3}$, which suggests that quantum effects will be important, and the classical estimate for the mass of the membrane-particle will not be reliable. In any case, as long as the typical size of the membrane-particle is much smaller than the Hubble scale (so that $T^{1/3}\gg H$), dimensional analysis forces a relation of the form
\begin{equation} E\sim g^\frac{\alpha}{4} T^{1/3},\label{Tm}\end{equation}
for some value of $\alpha$. For instance, using the classical electrostatic estimate for the energy of the membrane-particle coming from the electrostatic energy of a distribution with radius \eq{r4} gives $\alpha=16/3$. Assuming that the vacuum decays via flux discharge, it is natural to assume $\Delta V < V$, since the latter corresponds to full discharge of the flux. Stability under discharge then implies 
\begin{equation}
T^{4/3} \gtrsim\Delta V >V\,,
\end{equation}
and combining with equation \eq{Tm} leads to a relation of the form
\begin{equation} \frac{m^4}{V}\gtrsim g^{\alpha}\end{equation}
for some $\alpha$. This is precisely the form of the FL inequality \eq{FL0}, for $\alpha=2$. The classical value $\alpha=16/3$ produces a weaker bound; as we will see, in concrete examples where the mass can be computed precisely, such as the KPV setup discussed in Section \ref{sec:app}, corrections change the value of $\alpha$ to something $\geq2$. This establishes that a natural microscopic mechanism which is familiar from string compactifications, brane polarization, naturally leads to a relation of the FL type.

Another interesting aspect of this construction is that the charged states predicted in this way are part of a larger tower of particles which would be obtained by quantizing the membrane (similarly, but much more difficult, to what we do with the fundamental string in perturbative string theory \cite{deWit:1988wri,deWit:1988xki}). Thus, in this scenario, the charged particles satisfying FL are accompanied by neutral ones whose masses satisfy roughly the same relation $m^4\sim V$. We cannot help but notice that this is precisely the observed relation for neutrino masses \cite{ArkaniHamed:2007gg,Arnold:2010qz,Ooguri:2016pdq,Ibanez:2017kvh,Hamada:2017yji,Ibanez:2017oqr,Lust:2017wrl,Gonzalo:2018tpb,Gonzalo:2018dxi,Gonzalo:2021fma,Rudelius:2021oaz}, and although the picture outlined here is very sketchy, with many important details we have not addressed such as e.g. the lifetime of these neutral particles, we feel it deserves further study.  

Finally, we should remark that the mechanism outlined here, relying on the coupling \eq{mcop}, probably only applies in four dimensions or lower, since the estimate \eq{Tm} produces a different power of $T$ in higher dimensions, while \eq{FL0} remains the same in any dimension, and the agreement between the two is lost. This suggests that the ``membrane-particle'' mechanism for higher-dimensional de Sitter is specific to four(or less)-dimensional solutions with positive cosmological constant.

\section{FL and dimensional reduction}\label{sec:dimred}
Consistency of a Swampland bound under dimensional reduction can serve as both a stringent consistency check and a fruitful avenue to learn new information about Swampland conjectures \cite{Heidenreich:2015nta,Heidenreich:2019zkl,Rudelius:2021oaz}. We will now apply the same kind of logic to Festina Lente. 
To keep things concrete, we will consider dimensional reduction of the pure gravity sector of a theory in $(d+1)$-dimensions on a circle.

Our theory in $(d+1)$ dimensions is simply Einsteinian gravity with positive cosmological constant:
\be
S =\frac{1}{8\pi G_{d+1}} \int \sqrt{|g|}\left(R - 2\Lambda_{d+1}\right)\,.
\ee
Reduction on $S^1$ proceeds via the ansatz:
\be
ds^{2}_{d+1}=e^{-2\alpha\phi}ds^2_d + R_0^2 e^{2(d-2)\alpha\phi}(dy-A)^2.
\ee
Here $A_\mu$ is the Kaluza-Klein vector. There is a single modulus $\phi$, which parametrizes the size of the additional circle.  To canonically normalize the volume scalar $\phi$, we chose
\be
\alpha = -\frac{1}{\sqrt{2(d-1)(d-2)}}\,.
\ee
With this choice of $\alpha$, the physical size of the circle in $(d+1)$ dimensions is
\begin{equation} R=R_0\,e^{\kappa\phi,},\quad \kappa\equiv \sqrt{\frac{d-2}{2(d-1)}},\end{equation}
so that $\phi\rightarrow \infty$ is the decompactification limit.

The effective action in $d$ dimensions then reads \cite{Heidenreich:2015nta}
\begin{equation}S= \frac{1}{16\pi G_d} \int\sqrt{|g|}\Bigl( \mathcal{R}-\tfrac{1}{2}(\partial\phi)^2 -\tfrac{R_0^2}{2}e^{\gamma\phi} F_{\mu\nu}F^{\mu\nu} - 2\Lambda_{d+1}e^{-\delta\phi} -(16\pi G_d) V_{\text{Casimir}}(\phi)\Bigr),\label{e32}\end{equation}
where $ G_d=({2\pi R_0)^{-1}G_{d+1}}$ and
\begin{equation} \gamma=\sqrt{\frac{2(d-1)}{d-2}},\qquad \delta=\frac{\gamma}{d-1}=-2\alpha.\end{equation}
The KK photon above is not normalized canonically. The canonical normalization can be obtained by demanding that charge (KK momentum) is integer quantized. With this normalization, the right value of the gauge coupling in $d$ dimensions is \cite{Heidenreich:2015nta}
\begin{equation} g_{\text{KK}}^2= \frac{8\pi G_{d}}{R_0^2}e^{-\gamma\phi}.\label{gcop4}\end{equation}

In \eq{e32}, we have also included a Casimir energy term $V_{\text{Casimir}}(\phi)$, coming from a one-loop contribution of the matter fields in the UV theory . This takes the form \cite{Gonzalo:2021fma}
\begin{equation} V_{\text{Casimir}}(\phi)= \frac{\mathcal{C}}{R_0^{d}}\,e^{-\frac{d}{2}\gamma\phi},\end{equation}
where $\mathcal{C}$ is a function that counts the number of effectively massless bosonic vs. fermionic degrees of freedom at the scale set by the inverse KK radius. It is locally constant, and only changes when crossing mass thresholds of the parent $(d+1)$-dimensional theory. 

The combined potential for the radion $\phi$ may or may not have a minimum, depending on the details of the spectrum of the $(d+1)$-dimensional theory. For instance, the Standard Model reduced on a circle famously has a minimum where the Casimir energy due to the neutrinos can balance the contribution of the four-dimensional cosmological constant. A minimum with positive vacuum energy is only possible near a mass threshold; for the SM model on a circle, the neutrinos combine with the contribution of the 4d cosmological constant to produce such a minimum around $V_{d+1}$ \cite{ArkaniHamed:2007gg,Gonzalo:2021fma}.

Let us check what conditions are imposed by FL on this scenario, where the radion has been stabilized by the combined effect of Casimir and $(d+1)$-dimensional vacuum energies to some value $\phi_{\text{min}}$. In $d$ dimensions, we can apply the FL inequality to the Kaluza-Klein $U(1)$ so that \eq{FL0} becomes a bound on the mass of KK modes. The gauge coupling is\footnote{Reference \cite{Heidenreich:2015nta} defines the gauge coupling in such a way that the dilaton dependence is stripped away. In this paper we keep it, in line with \cite{Montero:2019ekk}.} \eq{gcop4} evaluated at $\phi_{\text{min}}$. Since the dilaton has been stabilized, we can ignore the first condition in \eq{FL0} and directly apply the bound on the spectrum of massive states. For a mode with KK charge $q$, this becomes
\begin{equation} m^4_{\text{KK}} \geq \frac{8\pi G_{d}}{R_0^2}\, q^2\, e^{-\gamma\phi_{\text{min}}} V.\label{f56}\end{equation}
In particular, we may apply this to the Kaluza-Klein modes of the  $(d+1)$-dimensional graviton itself, which are always present and have masses (in $d$-dimensional Planck units) 
\begin{equation}m^2_{\text{KK}}\sim \frac{n^2}{R_0^2}\, e^{-\gamma\phi_{\text{min}}}.\end{equation}
Combining with \eq{f56}, and setting $q=1$, we obtain a bound relating the stabilized value of the radion field and the vacuum energy $e^{-\gamma\phi}\gtrsim V$. 
This can be rewritten purely in terms of the physical KK scale in $d$-dimensional Planck units, as 
\begin{equation}M_{\text{KK}}\gtrsim V^{1/2}.\label{scalesep}\end{equation}
 Equation \eq{scalesep} is the main result of this Section. It says that the cut-off of the $d$-dimensional field theory must be above the $d$-dimensional Hubble scale. One could worry that this is just a self-consistency condition of the FL picture, or even of dimensional reduction; but the computation based on the Schwinger effect  works even when there is a whole tower of light particles, one just needs to consider the tower version of the conjecture we outlined in Section \ref{secFLtowerbound}\footnote{In this situation the tower bound provides only a very mild enhancement of charged particle production. The mass of the KK mode fields goes as $m_n\sim n M_{KK}$ while their charge goes as $q_n \sim n q$. The particle pair production is then enhanced as $\sum_{n=1}^\infty e^{-n M_{KK}^2 / q E}=1/(e^{M_{KK}^2 / q E}-1)$, which is only a very small enhancement when $m^2/qE>1$ as required to satisfy the ordinary FL bound. We note however that the presence of any other fields with mass lighter than the KK scale in the higher-dimensional theory, such as a photon, will further enhance the FL bound.}. We also note that the interesting case of $d+1=4$ must be excluded from the discussion since there are no charged Nariai black holes in three dimensions. With this caveat, we conclude that \eq{scalesep} is a constraint that should be imposed for consistency of a KK vacuum with positive vacuum energy, and that FL provides a modest lower bound on the size of the extra dimensions. 
 
 The FL inequality is not saturated in the real world. The electron, the lightest charged state, obeys the bound, but with a large margin. If we had a similar behavior for the KK modes, it would suggest that a perturbatively stable de Sitter KK vacuum in $d\geq 4$ is naturally scale-separated, with $M_{\text{KK}}\gtrsim V^{1/2}$.
This is to be contrasted to the behavior of known string compactifications to $AdS$, which tend to not have scale separation \cite{Gautason:2015tig}, and the behavior predicted by the AdS distance conjecture \cite{Lust:2019zwm}. This fits  with the idea that Minkowski vacua constitute a ``great divide'' with qualitatively different behavior on each side.

We also comment briefly on the case where the radion is unstabilized. Consider the asymptotic regime of large radius, $\phi\rightarrow\infty$. The Casimir energy contribution decays faster than that of the tree-level potential, and so without loss of generality, we can neglect its contribution. Since the field dependence of the gauge coupling and vacuum energy have opposite signs, there are no electric Nariai solutions in this limit. Magnetic solutions do exist, but not at parametrically large radius, due to the fact that the Nariai black hole has to fit in the cosmological horizon.  The details depend on the interplay of Casimir versus tree-level energy at large radius, and can potentially lead to constraints in the number of massless fields contributing to the vacuum energy. We will explore this in future work.

So far, we have discussed what happens for the KK $U(1)$. Another possibility is that one already has a gauge field in $d+1$ dimensions, which becomes a $U(1)$ gauge field in $d$ dimensions. We will now discuss briefly two such scenarios.

When there is an ordinary $U(1)$ gauge field in $(d+1)$ dimensions, the higher-dimensional FL bound is
\begin{equation} m^4\gtrsim g^2_{d+1} V_{d+1},\end{equation}
where $g_{d+1}^2$ is the $U(1)$ gauge coupling in $(d+1)$-dimensions. It is related to the $d$-dimensional gauge coupling $g_d$ as \cite{Heidenreich:2015nta}
\begin{equation} g_d^2= \frac{g_{d+1}^2}{2\pi R_0} e^{-2\alpha\phi},\end{equation}
where we have taken into account additional factors coming from the rescaling to Einstein frame. In $d$-dimensions, the FL inequality for the dimensionally reduced $U(1)$ is strongest for the KK modes. A canonically normalized scalar field of mass $m$ gets a mass $m\, e^{-\alpha\phi}$ in $d$ dimensions, owing to the conformal factor. As a result, the $d$-dimensional version of FL can be rearranged to
\begin{equation} m^4\gtrsim e^{2\alpha\phi} g^2_{d+1} \frac{V_{d}}{2\pi R_0}= g^2_{d+1} V_{d+1} \left( e^{2\alpha\phi}\frac{V_{d}}{2\pi R_0V_{d+1}}\right).\label{eom4}\end{equation}
This equation has interesting consequences.  The FL inequality can get stronger or lower after dimensional reduction, depending on the behavior of the term under parenthesis. If we demand that FL is automatically preserved under dimensional reduction, the constraint can only get weaker, and so
\begin{equation} e^{2\alpha\phi}V_{d} \lesssim V_{d+1}.\end{equation}
In the case where the $d$-dimensional vacuum is not stabilized, we have $V_d=(2\pi R_0) V_{d+1}e^{-\delta\phi}$, and since $\delta=-2\alpha$ the inequality is saturated: FL is preserved under dimensional reduction.

The situation is more interesting in a vacuum stabilized by Casimir energy. As mentioned above, a simple analysis of the Casimir + tree level potential shows that, for constant $\mathcal{C}$, one can only have $dS$ maxima or $AdS$ minima. As a result, stabilized KK vacua of positive vacuum energy can only appear near mass thresholds. The minimum can arise due to an interplay between bosons and fermions, or due to an interplay with the vacuum energy, as is the case for the Standard Model \cite{ArkaniHamed:2007gg,Gonzalo:2021fma}. We will just study this last case, where the minimum happens at
\begin{equation} R=R_0 e^{\kappa\phi}\sim V_{d+1}^{-\frac{1}{d+1}},\quad V_d\sim V_{d+1}^{\frac{d}{d+1}}.\end{equation}
Then, \eq{eom4} implies that $e^{(2\alpha+\kappa)\phi_{\text{min}}}>1$, or equivalently, $\phi_{\text{min}}>0$. FL is also automatically satisfied for the $U(1)$ in this case, but more generally, it might constrain the spectrum of light fields in the $(d+1)$-dimensional vacuum, simply from demanding that minima violating \eq{eom4} do not occur. It would be interesting to explore this in more detail although we must emphasize that, unfortunately, the phenomenologically most interesting case of $d+1=4$ is excluded from our analysis since Nariai black holes do not exist in $d=3$.

Finally, we will consider the case where the higher-dimensional $U(1)$ comes from a higher-dimensional gauge field. To keep the analysis short, we will only analyze a 2-form field in $(d+1)$ dimensions, but the expressions we derive are valid for a $(n+1)$-form in $(d+n)$ dimensions as well. We will start with a $(d+1)$-dimensional Lagrangian which, in addition to the cosmological and Einstein-Hilbert terms, also has a kinetic term for a 2-form $B$,
\begin{equation}\mathcal{L}_B= -\frac{1}{12 (g^B_{d+1})^2} H^2, \quad H=dB.\end{equation}
Dimensional reduction produces a $U(1)$ coming from the Wilson line of $B$ on $S^1$ \cite{Heidenreich:2015nta}. The resulting $U(1)$ gauge field in $d$ dimensions has gauge coupling
\begin{equation}\frac{1}{g_d^2}=\frac{1}{(g^B_{d+1})^2} e^{-2(d-3)\alpha\phi}.\end{equation}
The string charged under $B$, with tension $T$, can wrap the $S^1$, and becomes a particle of mass
\begin{equation}m=e^{(d-3)\alpha\phi}.\end{equation}
Assuming as well that we are applying the FL bound asymptotically, so that the potential is $V_{d+1} e^{-\delta\phi}$, we obtain the inequality
\begin{equation} T^4 e^{-2(d-4)\alpha\phi}\gtrsim g_{d+1}^2 V_{d+1}.\end{equation}
Interestingly, for $d=4$, the powers of $\phi$ drop out, and we recover an inequality for the tension directly in $(d+1)$-dimensional terms, or more generally, for the tension of an $n$-brane in $4+n$ dimensions. It is tempting to regard this as a generalization of the FL bound to these higher-dimensional cases. This is an interesting point to be explored in the future.

\section{Phenomenological applications}\label{sec:ph}
By far, the most exciting prospect of testing quantum gravity in quasi-de Sitter space is testing its phenomenological implications, since our own universe is believed to be one . Some of these were discussed in \cite{Montero:2019ekk}; we review this discussion and point out several new implications of FL which had not appeared before. 

\subsection{Standard Model and Higgs potential}
The most straightforward application of FL to the Standard Model is to check that \eq{FL0} is satisfied by the spectrum of electrically charged states \cite{Montero:2019ekk}. In particular, applying the bound to $W$ bosons one obtains a constraint
\begin{equation} v^2\gtrsim \frac{1}{g}M_PH= \frac{V^{1/2}}{g},\end{equation}
that partially explains the hierarchy between Planck and electroweak scales, given a positive vacuum energy. More precisely, \eq{FL0} can be recast as the statement that the cosmological constant in Planck units has to obey the bound
\begin{equation} \Lambda\lesssim \frac{m^4}{4\pi\alpha},\label{ler}\end{equation}
where $\alpha$ is the corresponding fine structure constant and $m$ is the mass of any electrically charged state. 

Although in the present section we are focused on the Higgs potential and the Standard Model, if FL is correct, equation \eq{ler} is a universal bound in models involving Einsteinian gravity and a positive cosmological constant. It explains partially the cosmological hierarchy problem, tying it to properties of the states of the effective field theory.
Going back to the real world, we may apply \eq{ler} to the lightest charged state, the electron. Then, the bound becomes (again in Planck units)
\begin{equation}\Lambda \lesssim 3\cdot 10^{-89},\end{equation}
which does a reasonably good job in bridging the 120 orders of magnitude between the Planck scale and the observed value of $\Lambda\sim10^{-120}$. 

We may also apply it to non-abelian gauge fields, like $SU(3)$, which remains unbroken. A non-abelian gauge theory automatically contains massless charged states, the gluons. Nariai black holes can be constructed by embedding the standard Nariai solution in the Cartan of the non-abelian gauge group, so massless non-abelian gauge fields are really in contradiction with \eq{FL0}. In other words, FL predicts that in a de Sitter background non-abelian gauge fields must confine or be spontaneously broken, at a scale above $H$. Explicitly,
\begin{equation} m_{\text{Gauge field}}\gtrsim H,\quad \text{or}\quad \Lambda_{\text{Confinement}}\gtrsim H\end{equation}
for non-abelian gauge fields. This is again satisfied in the real world, with $SU(2)$ being Higgsed, and $SU(3)$ confining. 

Similarly, the FL argument does not apply to massive vector fields whose mass is above the Hubble scale\footnote{More precisely, there can be no massless charged states whose mass is below $V^2 M_P^2$ where $V$ is the contribution coming from the dark energy. This is satisfied today, but becomes a strong constraint on the matter sector during inflation.}. The fact that FL is satisfied for the electromagnetic $U(1)$ in the real world is suggesting that the photon is exactly massless, since otherwise there would have been no reason to satisfy the FL bound. See \cite{Reece:2018zvv} for other Swampland arguments suggesting an exactly massless photon.

\subsection{Bounds on the Higgs potential}
\begin{figure}[!htb]
\begin{center}
\includegraphics[width=0.85\textwidth]{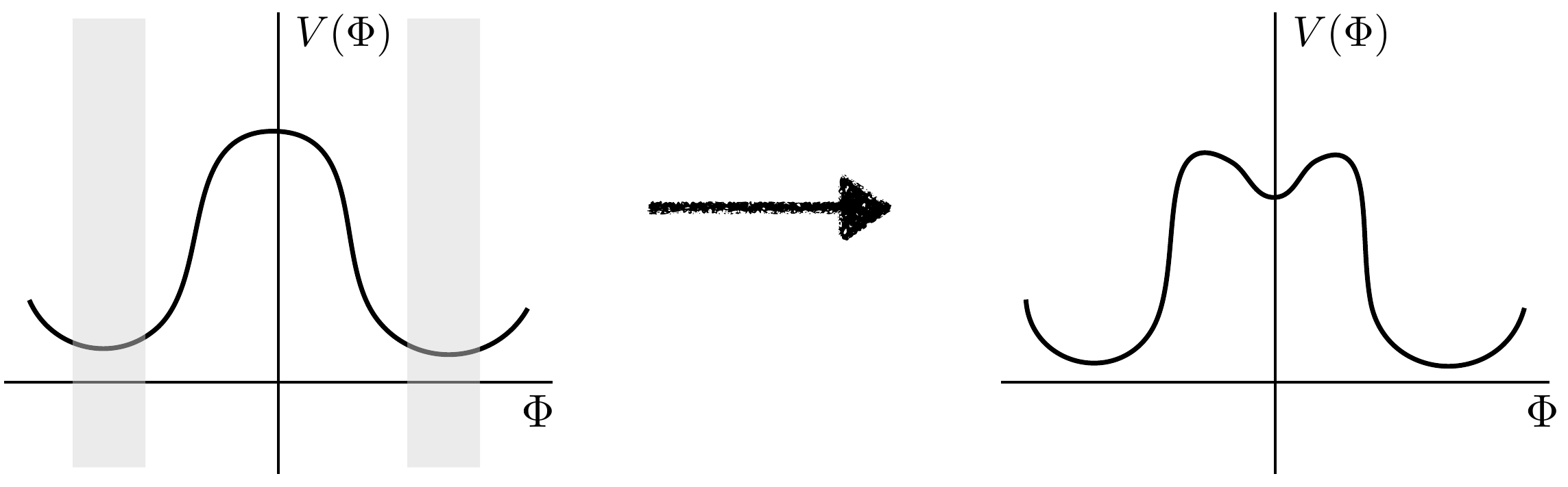}
\vspace{0.3cm}
\caption{\small{On the left, we show the usual shape of the ``Mexican hat'' Higgs potential, which arises from equation \eq{renpot}. However, only the region shaded in gray has been accessed experimentally. It is conceivable that the region near $\Phi\approx0$ has a different shape, for instance, that of the ``cowboy hat'' potential depicted in the right panel. As shown in the main text, such a scenario is incompatible with Festina Lente, unless extreme fine-tuning is carried out. }}\label{fig4}
\end{center}
\end{figure}

Festina Lente also leads to non-trivial constraints on the Higgs potential. LHC measurements have provided us with information about the Higgs mass and the quartic Higgs coupling, which fully determine a renormalizable potential \cite{Tanabashi:2018oca}:
\begin{equation} V(\Phi)= \mu^2 \Phi^\dagger \Phi+ \lambda ( \Phi^\dagger \Phi)^2.\label{renpot}\end{equation}
However, the measurements themselves do not tell us that the exact potential is of this form. While we expect significant corrections for large Higgs vevs, coming from loops of Standard Model and possibly new particles, we also do not have direct experimental evidence of the form of the Higgs potential around $\Phi=0$. In particular, one could imagine a situation like the one depicted in Figure \ref{fig4}, where non-renormalizable terms are added to \eq{renpot} in such a way that the Higgs potential develops a local minimum at the origin, and the potential looks more like a ``cowboy hat'' than the familiar ``mexican hat'' shape.

In the cowboy hat scenario, the electroweak symmetry is unbroken in the minimum at the top of the hat. Without fine-tuning, the vacuum energy is of order the Higgs vev,
\begin{equation} V_{\text{sym}}\sim (250\, GeV)^4,\end{equation}
and the electron and other hypercharged states remain classically massless, therefore violating FL. In other words FL predicts that we must be in a broken phase:  any long-lived minimum of the Higgs potential must break the electroweak symmetry. The confinement scale of the $SU(3)$ gauge fields is also way below the vacuum energy, unless additional fine-tuning is included. Thus, FL is incompatible with the ``cowboy hat'' scenario for the effective potential.

The simple analysis above applies to the effective quantum potential for the Higgs field in the IR, where all quantum effects have been integrated out. But it is also instructive to figure out what are the constraints applying to the UV Higgs potential, defined at the cut-off scale. To understand this, we must discuss the classical piece and quantum effects separately. This makes the full story subtler, as we will now explore\footnote{We are indebted to Matt Strassler for the argument in the main text.}, but we will still conclude that also the UV potential cannot have a local minimum.

 As explained e.g. in \cite{Hill:2002ap}, in absence of a symmetric minimum, the quark bilinears develop a condensate due to strong coupling effects in $SU(3)$, 
\begin{equation} \langle Q\bar{q}\rangle \sim \Lambda_{QCD}^{3}.\label{qbil}\end{equation}
Notice that the value of  $\Lambda_{QCD}$ could be different to that in our vacuum. In particular, the mere fact that the quark masses are set to zero will change $\Lambda_{QCD}$ by an $\mathcal{O}(1)$ factor; but there could be additional sources that change this number.

Due to \eq{qbil}, the masses of the W and Z bosons are of order $\Lambda_{QCD}$. To ensure compliance with FL in the non-abelian sector, we just need
\begin{equation}\Lambda_{QCD}\gtrsim H\sim \sqrt{V_{\text{sym}}},\label{ew3}\end{equation}
where $V_{\text{sym}}$ is the vacuum energy at the symmetric minimum. Unless extreme fine-tuning takes place, we would expect $V_{\text{sym}}$ to be of the order of the characteristic energy scale we observed in the Higgs potential, so of order hundreds of GeV. In this case, $H\sim 10^{-8}\, eV$ and \eq{ew3} is amply satisfied with the observed value of $\Lambda_{QCD}$.

The quark bilinear \eq{qbil} has the same quantum numbers as the Higgs field, so it triggers electroweak symmetry breaking $SU(2)\times U(1)_Y\rightarrow U(1)$. Let us call the local mass of the Higgs field as $\mu$, by analogy with \eq{renpot}; while in the usual renormalizable potential we have $\mu^2<0$, in the ``cowboy hat'' scenario we have $\mu^2>0$. Integrating out the Higgs field of mass $\mu^2$ will introduce non-renormalizable operators in the effective field theory, including couplings
\begin{equation} \frac{y_e y_q}{\mu^2} (L e_R) (Q\bar{q}),\end{equation}
where $y_e,y_q$ are the electron and quark Yukawas. Assuming the Yuakwas are unchanged at the symmetric point (an assumption that could also be challenged), the largest contribution will come from the top quark. Then  \eq{qbil} turns the higher dimension term into a mass term for the electron,
\begin{equation} m_e\sim y_e y_t |\frac{\Lambda_{QCD}^3}{\mu^2}.\label{melec}\end{equation}
This expression is valid when $\mu\gg \Lambda_{QCD}$; for $\mu \sim\Lambda_{QCD}$, the Higgs cannot be integrated out.  $QCD$ effects will generate a potential for it, and assuming the vacuum is not destabilized, it is reasonable to expect an effective mass of order $\Lambda_{QCD}$.
Imposing the FL bound then becomes a condition on $\mu^2$, the vacuum energy, and $\Lambda_{QCD}$,
\begin{equation} \mu^8\leq y_e^4 y_t^4\frac{\Lambda_{QCD}^{12}}{g V_{\text{sym}}}.\end{equation}
Substituting Standard Model values on the right hand side, as well as $V_{\text{sym}}$ of order a hundred $GeV$, the above becomes 
\begin{equation}\mu\lesssim 10^{-4}\, eV.\end{equation}
This is outside of the regime of validity of \eq{melec}, so the effective mass for the electron will be of order $\Lambda_{QCD}$. This is still below the vacuum energy at a few hundred GeV.

Thus, assuming all other parameters remain equal, Festina Lente puts an upper bound on the curvature of the ``cowboy hat'' at the top of the Higgs potential, in either its IR or UV versions. It is so strong that, in practice, we go back to the usual Mexican hat potential. So in a sense, Festina Lente predicts that the shape of the potential is the usual Mexican hat one, but we should stress the assumptions that went into this for the UV discussion. First, we assumed that all the relevant SM parameters do not change significantly between our vacuum and the symmetric one; we also assumed that no new sectors/field directions arise that might destabilize the potential, and we assumed a ``natural'' potential which is not extremely fine-tuned. More generally, Festina Lente implies that there is no natural (without large fine-tuning) nearby minimum to the SM vacuum, for $\langle\Phi\rangle\lesssim100\, GeV$. Some of the constraints obtained in this Section are reminiscent of those in \cite{Gonzalo:2018dxi}, which places constraints on putative SM vacua with unbroken electroweak symmetry by demanding that they do not lead to non-supersymmetric 3d AdS vacua, in order to comply with \cite{Ooguri:2016pdq}. In particular, \cite{Gonzalo:2018dxi} found that pions have to be massive enough in order to avoid the unwanted 3d vacua, in line with what FL demands.

\subsection{Neutrino physics}
The observed vacuum energy is tantalizingly close to the neutrino mass scale, and it is natural to look for an explanation of this coincidence \cite{ArkaniHamed:2007gg,Arnold:2010qz,Ooguri:2016pdq,Ibanez:2017kvh,Hamada:2017yji,Ibanez:2017oqr,Lust:2017wrl,Gonzalo:2018tpb,Gonzalo:2018dxi,Gonzalo:2021fma,Rudelius:2021oaz}. As observed in \cite{Montero:2019ekk}, there is a straightforward connection to FL, since the lower bound to the mass of charged states is roughly $V^{1/4}$. Unfortunately, one cannot apply the FL bound directly to the neutrinos, since they do not have electric charge in the Standard Model. We now comment on a couple of ways to connect the neutrino to FL that fail, and one that partially succeeds. 

Our first attempt starts with the observation that neutrinos do not have electric charge, but they do have magnetic dipole moments. One may wonder whether there is a bound involving neutrinos and magnetic Nariai black holes. Recall that an electric Nariai black hole has an electric field $ E\sim g M_P H$. For a magnetic black hole, the powers of $g$ work out to give the same magnetic field:
\begin{equation} B\sim g M_P H.\end{equation}
This is in accordance with electric-magnetic duality, which swaps $E\leftrightarrow B/g^2$  and $g\leftrightarrow 1/g$ when charges are involved, to preserve quantization properties.

A magnetic Nariai black hole will discharge slowly by pair-production of monopoles, which are quite heavy. While electrically charged particles are not produced, they can still lead to instabilities, coming from their magnetic moment. A particle of magnetic moment $\mu$ properly oriented gets a contribution to its energy given by $-\mu B$. If these particles have mass $m$, one might expect some sort of instability whenever
\begin{equation} \mu B > m,\label{wee4}\end{equation}
so that is energetically favourable to pair-produce this particles and ``fill the sea''.

Avoiding \eq{wee4} puts a bound on the magnetic moment
\begin{equation} \mu \lesssim \frac{m}{g M_p H}\label{www}\end{equation}
for any particle. As a consistency check, notice that for a Dirac fermion, the magnetic dipole moment is a kinematic property, and it is always given by $\mu = g/m$ plus small corrections, where $m$ is the particle mass and $g$ is the coupling. Plugging back in \eq{www}, we recover the FL bound
\begin{equation} m^2\gtrsim g^2 M_P H.\end{equation}
In contrast to FL, however, \eq{www} can also be applied to neutral particles, as long as they have a nonzero magnetic moment. As explained in \cite{Broggini:2012df}, a Dirac neutrino gets an effective magnetic moment in the SM augmented with right-handed neutrinos, from loop factors, roughly
\begin{equation}\mu_\nu^{\text{SM}} \sim  \frac{m_\nu}{M_W^2}\sim \frac{1}{10^{14}\, \text{GeV}}\end{equation}
Imposing \eq{www}, we get that the masses drop out and the consistency condition is simply
\begin{equation} M_W^2> g M_P H,\end{equation}
which is correct, but which is already a consequence of ordinary electric FL \cite{Montero:2019ekk}. If there is new physics at some scale $M$ below $10^{14}\,\text{GeV}$, this could generate a magnetic moment of $1/M$. Applying \eq{www} we get the more interesting
\begin{equation} m\geq \frac{ g M_P H}{M}.\end{equation}
This is automatically satisfied again, since we know that $M\gg m$. We note in passing that the electric and magnetic fields of Nariai black holes in our own universe are of order
\begin{equation} E_{\text{Nariai}}\sim 100\, V,\quad B_{\text{Nariai}}\sim 0.01 \, G.\end{equation}
These are comparable to electric and magnetic fields in the surface of the Earth. Thus, we are very unlikely find any new constraints on particle properties, such as e.g. neutron electric dipole moment, just from studying their decay -- we would have seen them long ago. 

Thus, our first attempt at explaining the neutrino mass via FL does not work. For our second attempt, we observe that although the neutrino has no electric charge, it is charged under electroweak interactions, and carries $B-L$ charge. However, if $B-L$ is gauged, its coupling is experimentally constrained to be very weak, of order $g_{B-L}\sim 10^{-24}$ \cite{Craig:2019fdy} (and from the ordinary magnetic WGC, we get $g_{B-L}\sim 10^{-28}$, see \cite{Craig:2019fdy}) . Therefore, although FL would provide a lower bound in this case, we would lose the straightforward connection to the vacuum energy. Another, closely related possibility is that $B-L$ is spontaneously broken in our vacuum, and that some version of FL applies to massive $U(1)$ fields. The argument involving Nariai black holes in \cite{Montero:2019ekk} only works for vector fields whose mass is of order Hubble or smaller, since otherwise the Nariai black holes will not be long-lived. And our discussion about anti-branes in the KS throat showed that one can have very light states charged under a spontaneously broken symmetry in the non-compact scenario, where gravity is decoupled.

Finally, we describe our last, and partially successful attempt.  As emphasized near the end of Section \ref{sec:rev}, in the ``membrane-particle scenario'', FL is satisfied because the domain walls that trigger vacuum decay have worldvolume degrees of freedom that can induce electric charge, and these states obey a relation of the FL form. In this scenario, there are additional, neutral states, coming from neutral membrane ``blobs'',  which also satisfy a relation of the form 
\begin{equation} m^4 \sim V.\end{equation}
It would be natural to identify the neutrinos with these degrees of freedom. The right-handed neutrinos could then naturally be the lowest step in a tower of states with $m\sim V^{1/4}$, leading to a scenario like the one proposed in \cite{Dienes:1998sb} where the neutrino mass eigenstates are nontrivial linear combinations of the left-handed neutrinos and the tower of right-handed ones. 

If this scenario is correct, it would provide a microscopic explanation of the tantalizing match between the neutrino mass scale and the vacuum energy, an area that has been the subject of intense recent research \cite{ArkaniHamed:2007gg,Arnold:2010qz,Ooguri:2016pdq,Ibanez:2017kvh,Hamada:2017yji,Ibanez:2017oqr,Lust:2017wrl,Gonzalo:2018tpb,Gonzalo:2018dxi,Gonzalo:2021fma,Rudelius:2021oaz}. We should emphasize that the analysis here is qualitative and that any quantitative improvement would need to address a number of pressing questions such as producing the right neutrino mass matrix, suppressing appearance of effective non-unitarity.

\subsection{Supergravity FI terms}
When it comes to de Sitter model building lots of work has been done both top-down and bottom-up. In this regard, swampland bounds are useful; For top-down research their violation serves as a warning sign that something is not under control in the construction. In bottom-up research they similarly provide warning signs that an eventual embedding of the suggested EFT in quantum gravity will fail. In neither of the two cases does a non-violation of Swampland bounds imply the model is consistent. 

The dS top-down attempts suffer from not being entirely top-down \cite{Danielsson:2018ztv}. The bottom-up models are in worse shape since there is no a priori reason why any 4D supergravity model with meta-stable dS can be lifted to string theory. 
It is  surprising how constraining it is to write supergravity theories with meta-stable dS vacua (see for instance \cite{Covi:2008ea}). The current state of affairs is that there is no example in 4D of meta-stable dS in theories with more than 8 supercharges. In theories with 8 supercharges ($N=2$) examples have been found and were hard to come by.  A popular idea has been that more supersymmetry of the Lagrangian increases the chances for it to descend from string theory.

In theories with 4 supercharges ($N=1$) there is a common lore that FI terms are the easiest road to meta-stable dS as well. The point we make here is that the bottom-up models typically violate the FL bound together with either the magnetic weak gravity conjecture or the no-global symmetry conjecture. This, once again, shows how Swampland bounds form a tight web. We start with $N=1$ supergravity and then discuss $N=2$. 

In $N=1$ SUGRA one gauges the R-symmetry to get constant (ie field independent) Fayet Iliopolous (FI) terms, which help in providing meta-stable dS vacua since they contribute the following term to the potential\footnote{We follow reference \cite{Catino:2011mu} which contains a particular useful discussion about FI terms.}:
\be
V_D = \tfrac{1}{2}D_a D^a\qquad D_a = i K_i X^i_a +\xi_a\,,
\ee
where $X_a = X^i_a\partial_{z_i}$ is the holomorphic Killing vector on the scalar manifold that is being gauged.  Note that indices $a$ are raised with the inverse gauge kinetic function. The constants $\xi_a$ are associated to gaugings of the R-symmetry. This gauge symmetry can be spontaneously broken and if that is the case everywhere in field space we can redefine this part of the D-term away in terms of an F-term by a Kahler gauge transformation. In what follows we have in mind the situation where this is not the case.

The gravitino is a Majorana field and a gauge-invariant mass term for it does not exist. But since the gravitino is charged under the R-symmetry, all these dS models with gauged R-symmetry violate the FL bound:
\be
\text{Constant FI terms} \quad \rightarrow \quad  \text{violation of}\quad  \eqref{FL0}\,.
\ee
Interestingly, this affects a substantial class of models in the supergravity literature and one can verify that the same models violate the magnetic WGC \cite{Cribiori:2020wch, Cribiori:2020use} as well. So a violation of the magnetic WGC seems to play together with a violation of the FL bound.

There is a deeper connection with the Swampland possibly explaining these issues; Seiberg and Komargodski argued that field independent FI terms are in the Swampland \cite{Komargodski:2009pc} because they require the theory to have a global ungauged symmetry:
\be
\text{Constant FI terms} \quad \rightarrow \quad  \text{global symmetry}.
\ee

The argument in \cite{Komargodski:2009pc} shows that there must be an exact global symmetry of the supergravity lagrangian, although it is not clear it cannot be broken by non-perturbative effects which do not have a lagrangian description. Here, we see that these considerations align nicely with FL. 

Concerning $N=2$ supergravity, a common recipe is that one needs at least the analogue of FI terms and non-compact gaugings \cite{Fre:2002pd}. Non-compact gaugings interestingly are claimed to be in the Swampland based on arguments not related to cosmology \cite{Banks:2010zn, Heidenreich:2021tna}. More recently, however, examples with compact gaugings were found  in \cite{Catino:2013syn} and \cite{Cribiori:2020use}. 
Both models rely on gauging the R-symmetry and have charged gravitinos. The first model has hypermultiplets and the second does not. For the model without hypers, reference \cite{Cribiori:2020use} computed that there is always a massless combination of gravitinos at the dS critical point and they hence violate FL. They also violate the recently proposed Gravitino Distance Conjecture \cite{Cribiori:2021gbf,Castellano:2021yye}, since these points would be at finite distance. To date, there is no proof that this will always happen since a priori one can write down gauge invariant mass terms for gravitinos in $N=2$ supergravity, but so far all examples have a massless charged gravitino. The second class of models \cite{Catino:2013qta} unavoidably feature light gravitinos for meta-stable dS vacua, with a mass of the order of the Hubble scale. Even more, \cite{DallAgata:2021nnr} have found evidence that the FL bound tends to be violated more generally.  

\section{A speculation: The decoupling limit of Festina Lente?}\label{sec:evi}
So far, we have focused in applying the logic and tools that have been useful in the past to learn about other Swampland constraints, notably the WGC \cite{ArkaniHamed:2006dz}. We now turn to a discussion that is particular to FL: the physics of the limit where gravity is decouped, $M_P\rightarrow\infty$.
The FL inequality \eq{FL0}, as written originally in 4D in \cite{Montero:2019ekk} 
\be \label{FLold}
m^2> gqM_pH \qquad \text{(for all charged particles in the theory)}\,
\ee
has a feature which makes it unlike any other Swampland constraint: the naive $M_P\rightarrow \infty$ at constant $H$ seems to forbid any charged states. As pointed out in \cite{Montero:2019ekk}, this reasoning is too quick, since we do not know how $H$ behaves as gravity is decoupled. We do not even know that the decoupling limit makes sense: One should expect de Sitter solutions to correspond to isolated points in the landscape \cite{Ooguri:2018wrx}, far away from asymptotic limits. Nevertheless, \eq{FLold} can be recast as in (\ref{FL0})
\be 
m^4> g^2q^2V \qquad \text{(for all charged particles in the theory)}\, \label{eew}
\ee
which becomes a nontrivial statement even when gravity is decoupled\footnote{Notice that $V$ is actually the same whether we work in Einstein or string frames, which makes \eq{eew} well-defined (FL was argued for in the Einstein frame). 
}.  This is to be contrasted with the WGC, which in the limit $M_P\rightarrow\infty$ becomes the trivial statement (as long as $g$ is kept constant)
\begin{equation}m\leq  g\cdot \infty. \end{equation} 
 
To analyze the FL bound in the limit of decoupling gravity, and thus non-compact models in string theory, we need to address a few subtleties first: \begin{itemize}
\item We need to define the vacuum energy $V$ in situations without gravity, and there is no canonical way to do this. One could even entertain the possibility of setting $V=0$ always, in which case FL would always be satisfied. However, in models where supersymmetry is broken spontaneously it is natural to set $V=0$ for the supersymmetric minimum. The vacuum energy $V$ is then uniquely defined for any metastable supersymmetry-breaking state over this minimum. This is the only case we consider. 
\item One can trivially violate \eq{FL0} in field theory simply by having two completely decoupled, free sectors, one of which breaks supersymmetry and another one which contains the gauge fields/charged matter. We will restrict FL to theories where all sectors are coupled to each other (there is a single stress-energy tensor) below the cut-off scale. More precisely, we demand that all couplings are $\mathcal{O}(1)$ in units of the cut-off at the cut-off scale, in order to discard situations with two almost-decoupled sectors. 
\item Finally, we will only consider theories with a single $U(1)$ gauge field at low energies. In general, when coupling a quantum field theory to gravity, relevant couplings become the vevs of dynamical fields. In non-supersymmetric setups, these can have runaway potentials, leading to instabilities and a violation of the inequalities in \eq{FL0}. What we have found ``experimentally'' in string configurations, detailed below, is that while setups with a single $U(1)$ field do not necessarily lead to such deformations, multi-$U(1)$ setups can have them. 
\end{itemize}

We will now see that, with the above caveats, \eq{FL0} is satisfied in a number of field and string theory examples.  We will also see explicitly that \eq{FL0} fails to hold in situations with more than one $U(1)$ in the deep IR. We will separate the discussion in two cases: pure quantum field theories without a known stringy embedding, and QFT's obtained explicitly from 10-dimensional non-compact string backgrounds.

To be clear, we are \emph{not} proposing \eq{FL0} as a constraint on general quantum field theories. Although there may well be a formulation of FL that holds in general QFT's, we do not have enough evidence, nor a purely field-theoretic motivation for doing so. In particular, it is already clear that FL only applies to metastable vacua of QFT's that will not be destabilized significantly by coupling to gravity. As explained above, in examples we have found that this is a problem whenever one has more than one $U(1)$ field at low energies, but it is unclear how general this is. Thus, the purpose of this Section is to collect some evidence for the bound in the decoupling limit, that we find somewhat nontrivial.

\subsection{FL in pure QFT?}
Equation \eq{FL0} is automatically satisfied in any supersymmetric vacuum. To find examples where the conjecture becomes nontrivial, we need to consider supersymmetric QFT's with metastable non-supersymmetric vacua. The canonical examples are the vacua of ISS in \cite{Intriligator:2006dd}. This reference showed that $\mathcal{N}=1$ SQCD with massive flavors exhibits dynamical supersymmetry breaking. For $N_f$ flavors and $N_c$ colors, the IR dynamics is controlled by a theory involving the following chiral fields; $\Phi_{ij}$, $\varphi_c^i$ and $\tilde{\varphi}_c^i$ with $i,j=1,\ldots N_f$ and $c=1,\ldots N_f-N_c$, and with a tree-level superpotential
\begin{equation} W=h\text{Tr} (\varphi\, \Phi\, \tilde{\varphi})- h\mu^2\text{Tr}(\Phi)\,,\end{equation}
where $h$ is a dimensionless coupling and $\mu$ a mass scale.

It is impossible to set the F-terms of this model to zero simultaneously, leading to supersymmetry-breaking vacua\footnote{Supersymmetric vacua exist upon the inclusion of quantum effects.}. The resulting low-energy theory does not have $U(1)$ gauge fields, and so \eq{FL0} is satisfied automatically. 
The theory has an exact, non-anomalous global symmetry, baryon number $B$ in \cite{Intriligator:2006dd}, which can be weakly gauged without affecting the IR dynamics of the strongly coupled sector. $B$ is spontaneously broken in the non-supersymmetric vacuum; the corresponding gauge field would become massive, precluding application of \eq{FL0}.  

 But on the other hand, the theory also has an exact, non-anomalous global symmetry, baryon number $B'$ in \cite{Intriligator:2006dd}, which can be weakly gauged without affecting the IR dynamics of the strongly coupled sector\footnote{Section 2 of \cite{Intriligator:2006dd} discussed a non-anomalous exact baryonic symmetry $B$, which is spontaneously broken in the metastable non-supersymmetric vacuum. In that vacuum however there is a second baryonic symmetry $B'$, obtained by combining $B$ with a Cartan generator of the $SU(N_f)$ flavor symmetry. This symmetry remains unbroken, and is the one we gauge here.}. The vacuum energy, $V$, is given up to $\mathcal{O}(1)$ factors as
\begin{equation} V\sim \Lambda^2m^2,\end{equation}
where $m$ is the typical scale of the quark masses, and $\Lambda$ is the strong coupling scale of the theory. The analysis in \cite{Intriligator:2006dd} is valid for $m\ll \Lambda$. The baryons have a mass $\sim \Lambda$, and therefore \eq{FL0} is satisfied for any gauge coupling $g\lesssim1$. 

Notice that the theory described above has a single $U(1)$ gauge field at low energies, and hence satisfies our criteria. There is a simple variation of the above story, where one just simply adds a completely decoupled sector including a $U(1)$ gauge field and a light or massless charged multiplet. \eq{FL0} is clearly violated in this case, but this is against the second point in our list above.

It would be very interesting to explore further evidence for FL in other proposals of supersymmetric QFT's with nonsupersymmetric vacua, such as those of \cite{Franco:2006es,GarciaEtxebarria:2006rw,GarciaEtxebarria:2007vh,Argurio:2006ny,Argurio:2007qk} involving fractional $D3$ branes. For instance, in the model described in \cite{Franco:2006es}, involving fractional branes in the cone over $dP_1$ and $D7$ branes that stabilize the runaway of the potential, the quiver gauge theory has three gauge factors, one of which confines, and the other two are fully Higgsed. Thus, FL is satisfied trivially, in the sense that all non-abelian gauge groups are broken or gauged. It would be very interesting to explore this question in further detail. 

\subsection{Anti-branes in the Klebanov-Strassler throat}
Consider the Klebanov-Strassler throat \cite{Klebanov:2000hb}. This is a supersymmetric,  non-compact background of 10-dimensional supergravity with $H_3,F_3$ fluxes turned on. The 10d metric (in string frame) can be written as
\begin{equation} \label{warpedCY}
ds^2 =  e^{2A}ds_4^2+ e^{-2A} ds^2_6
\end{equation}
$ds^2_6$ is the CY metric and $e^{2A}$ the warpfactor and its inverse the CY conformal factor. The tip of the throat is an $S^3$ (also known as the A-cycle) that is supported by $M$ units of flux going through it and its radius is of the order $\sqrt{g_sM}$ such that $Vol_A\sim (g_sM)^{3/2}$.  

Because of supersymmetry, the four-dimensional vacuum energy vanishes exactly. Putting $\overline{D3}$ branes on this background breaks supersymmetry; this is the Kachru-Pearson-Verlinde setup (KPV) \cite{Kachru:2002gs}. For $p$ anti-D3 branes with $p/M$ small enough this background was suggested to the dual dynamical SUSY breaking in the $SU(2M-p)\times SU(M-p)$ gauge theory at the end of the Seiberg  cascade.

The $U(1)$ gauge field that we focus on will be the one coming from the anti-D3 worldvolume. For a single anti-D3 (the case of interest) we have
\begin{equation} S_{\overline{D3}}\supset -\frac{1}{4\pi e^{\phi}}\int \vert F-B_2\vert^2.\end{equation}
In the above action we have set the background value of the RR axion $C_0$ to zero but included the 10d $B_2$ field to which the D3-brane couples to. Physically, this coupling ensures that charged states are endpoints of strings\footnote{The 4d zero mode of the $B_2$ field is projected out by the orientifolds in compact models.} 

The computations in \cite{Kachru:2002gs} were done instead with the S-dual worldvolume theory where $B_2$ is replaced with $C_2$ implying D1 strings attach to this object instead; However, as explained in \cite{Kachru:2002gs} this was merely a computational trick to get a handle on the calculation. In reality the regime of the theory is not in the magnetic phase but somewhere in between and both fundamental strings and D1 strings can attach. Below we consider indeed both magnetic and electric states. Although the computations by KPV relied on the S-dual action in the ``wrong regime", the predictions of the model stood a highly non-trivial test when identical results were rederived in the blackfold approach that is valid in the electric regime \cite{Armas:2018rsy}. We therefore take some confidence in the KPV results and rely on them from here onwards.  

The value of the vacuum energy is of the order
\be
V \approx 2pg_s^{-1}e^{4A}\,,
\ee
in string units.  The 4d gauge coupling depends on the value of the dilaton at the tip but since the dilaton is roughly constant, we can take it to be the zero mode and hence $g^2= 4\pi g_s$. We can now check the validity of \eq{FL0} for the states charged under this $U(1)$; 
 
\begin{figure}[ht]
 	\centering
 	\includegraphics[width=0.25\textwidth]{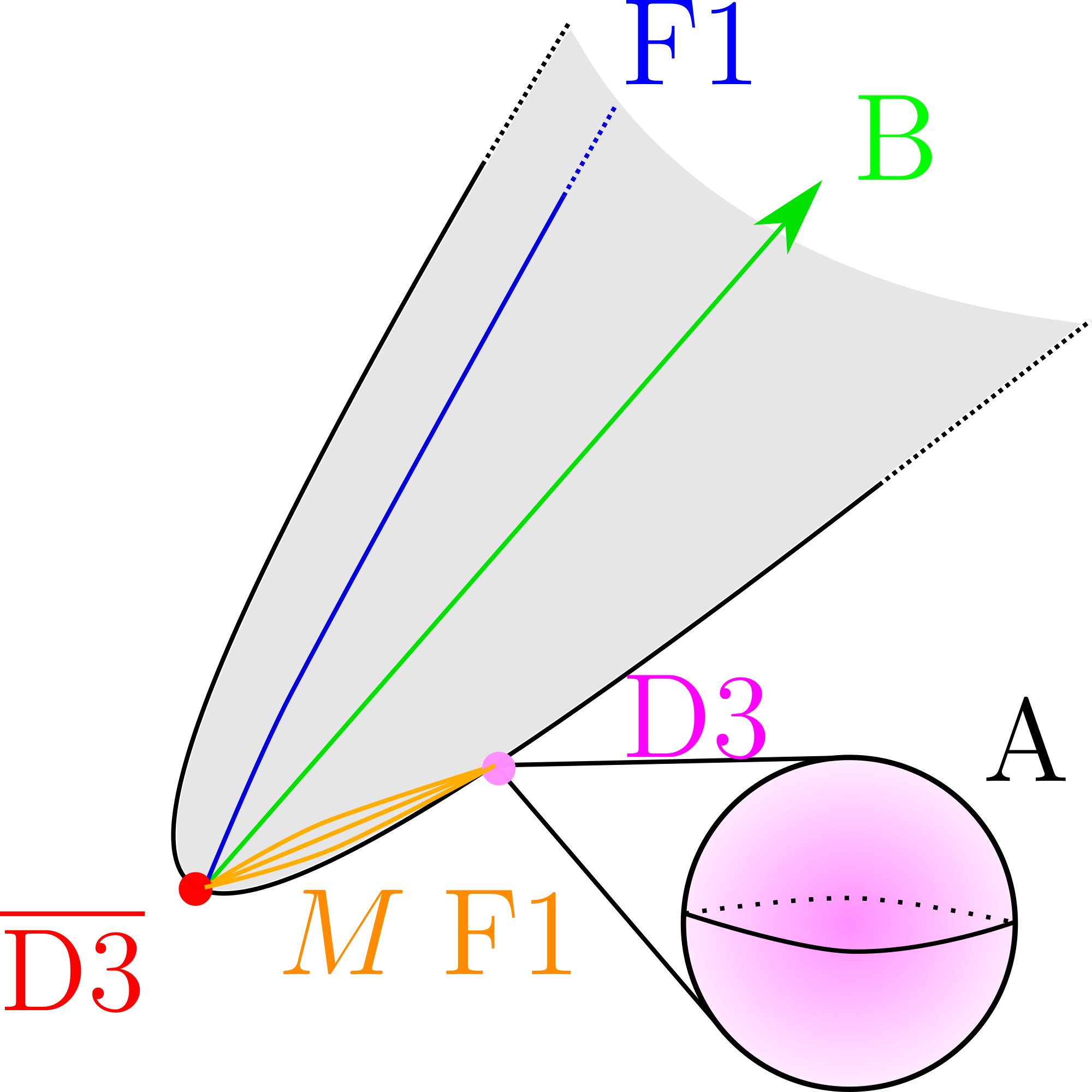}
 	\caption{The bottom of the KS throat in which every point has a finite $S^3$ attached to it due to the A-cycle. The anti-D3 is pointlike and has M stretched string attached to the D3 wrapping the full A-cycle (which is instead pointlike in 4D.) Those $M$ strings are depicted in orange. We also consider strings that start at the anti-D3 and leave the throat to end somewhere on a D7 in the bulk. Such an example string is depicted in blue.}
 	\label{fig: throat2}
 \end{figure}

 We will consider the following states:
 \begin{enumerate}
 	\item Strings that are attached to the anti-D3 branes and leave the throat (towards the D7s in the bulk). They carry unit electric charge and are depicted in blue in figure \ref{fig: throat2}. 
 	\item D3 branes wrapped on the $S^3$ at the tip of the throat (depicted as the red circle in figure \ref{fig: throat2}). Such D3 branes are pierced by $F_3$ flux with quantum $M$ and hence emit $M$ fundamental strings (depicted in orange) that can attach to the anti-D3. Therefore the wrapped D3 brane is an electric state of charge $M$. 
 \end{enumerate}

In non-compact throats there is no other point for the strings leaving the throat to end and the resulting states are infinitely massive, and satisfy FL in a trivial way. We will reconsider them in the next section when we discuss compact models.

We hence turn to the ``baryon particle" made from a $D3$ brane wrapped on the $S^3$ at the tip of the throat. The FL bound gives:
\begin{equation} m^4=e^{4A} T_{D3}^4(\text{Vol}_A)^4\gtrsim p M^2 g_s T_{D3} e^{4A}\qquad \rightarrow \qquad g_s^{2} M^4\gg p,\label{wee2} \end{equation}
since $T_{D3}=1/g_s$ and $\text{Vol}_A\sim (g_s M)^{3/2}$ (in string units) and we ignored factors of $\pi$. 
In short, \eq{FL0} becomes the constraint \eq{wee2} on the parameters of the KS throat. A consistency condition necessary for the validity of the supergravity description is that $\sqrt{ g_s M}\gg1$, and since $M\geq1$, we find that \eq{wee2} is automatically satisfied in the regime of validity of supergravity. Furthermore stability requires $p/M$ to be small enough (smaller than $0.08$ at least.). So the FL bound for the baryon is essentially guaranteed by stability of the anti-brane. Once coupled to gravity via a compact bulk, we gain the extra condition for meta-stability \cite{Bena:2018fqc}: $\sqrt{g_s} M> 6.8\sqrt{p}$.

We can also  consider keeping  $g_s$ finite and set $M\rightarrow0$ instead. In this case, before adding $\overline{D3}$ branes, the supersymmetry is enhanced to $\mathcal{N}=2$, and adding the $\overline{D3}$ just breaks to $\mathcal{N}=1$. Being a supersymmetric background, FL is satisfied by trivial reasons. In any case, notice that the (now massless) wrapped $D3$ branes are no longer charged under the $\overline{D3}$ worldvolume $U(1)$ gauge fields.

When $p>1$ there is a nontrivial vacuum manifold, and FL should be satisfied at the minimum of the potential. This does not affect too much our naive above analysis proving that the wrapped $D3$'s always satisfy FL. But we need to check that the $p$ $D3$ branes separate dynamically, as for instance argued for in \cite{Bena:2014jaa}. The $p$ antibranes will polarize into an NS5 brane which can be understood as a Higgsing of the worldvolume $U(p)$ gauge fields down to the center of mass $U(1)$. This is matched by the low-energy theory of the type IIB NS5 brane, which includes a $U(1)$ gauge field at low energies \cite{Kim:2015gha}. As described in \cite{Polchinski:2015bea}, the vacuum energy receives only small corrections for $p\ll M$, so the FL inequality in the form \eq{wee2} remains valid. 

It is important to remark that the asymptotic version of FL \eq{FL0} really requires a massless $U(1)$ to apply. At finite $M_p$ one could argue that the mass of the vector should at least be lighter than the Hubble scale. But in the decoupling limit this should turn into the requirement it is exactly massless. The KPV setup provides an explicit example of this. Consider the case of two antibranes ($p=2$). The antibranes polarize to a sphere of radius \cite{Kachru:2002gs}
\begin{equation} R^2 \approx 2\pi^2 \text{Tr}[\Phi^2]\approx 12\pi^2 \frac{g_s }{M}.\end{equation}
This is the radius in warped down string units.
Since this is the vev of the field ($\text{Tr}[\Phi^2]$) that gets a mass to describe the polarization, the above scale also the mass of the massive gauge bosons in string units. The FL inequality would then become
\begin{equation} g^2_s/M^2> 4\pi ,\end{equation}
which is just not true in the large $M$ limit at fixed $g_s$. In other words, the mass of the $W$ bosons can get much below $V^{1/4}$, although it is always above $V^{1/2}$ in Planck units.

\subsection{Confining gauge theory from \texorpdfstring{$D5$}{D5} branes}
We explained earlier how \eq{FL0} required that the deep IR configuration of the metastable supersymmetry-breaking state had a single $U(1)$ at low energies. The KPV setup we just described has this property; we will now show that it is actually required in stringy setups as well, by studying the configuration discussed in \cite{Aganagic:2006ex}. The setup is a resolved conifold, where the $S^2$ has finite size. One can wrap $N$ $D5$ branes around it, resulting in a 4d $\mathcal{N}=1$ theory without a moduli space. As discussed in \cite{Aganagic:2006ex}, the $SU(N)\subset U(N)$ of the theory confines in the IR, and the $U(1)\subset U(N)$ remains massless, with a gauge coupling equal to the bare UV coupling of the theory divided by $N$, coming from the trace in $U(N)$,
\begin{equation} g^2_{U(1)}= \frac{g_{YM}^2}{N}.\end{equation}
This is a supersymmetric background, so FL is satisfied.  A more interesting situation arises when the size of the $S^2$ varies from point to point, and there is more than one stack of $D5$ branes. Consider two stacks, sitting on nearby $S^2$ separated by a distance $\Delta$, by a potential barrier of height $g \Delta^3$. This can be engineered by considering a non-compact CY given by the hypersurface
\begin{equation} uv= y^2+ W'(x)^2\,\subset \mathbb{C}^4\,,\end{equation}
where
\begin{equation} W'(x)= gx (x-\Delta).\end{equation}
 We will look at the vacuum where we have a single $D5$ branes at $x=0$ and a single $\overline{D5}$ branes at $x=\Delta$, and consider FL for the antidiagonal $U(1)$.  There are charged fields coming from  bifundamental strings, corresponding to W bosons and a bifundamental Higgs field. The mass of this field is controlled by $\Delta$, and for $\Delta\lesssim1$ it is a tachyon -- the usual open string tachyon of a brane-antibrane pair \cite{Aganagic:2006ex}--. The resulting vacuum is unstable, and \eq{FL0} does not apply. But as we increase $\Delta$, the tachyon becomes massive, and in particular there is a phase transition at which it is exactly massless. At this point, or for $\Delta$ slightly larger, the former tachyon is massive, but very light, while the vacuum energy is given by the tension of the brane-antibrane pair,
\begin{equation} g_{YM}^2V= 16\pi\, - \frac{4}{\pi}g_{YM}^2 \log \left(\frac{\Lambda_0}{\Delta }\right),\label{vacd5}\end{equation}
where the second term is a one-loop correction that can be made arbitrarily small by going to weak coupling, and $\Lambda_0$ an UV cut-off scale. It is clear that this theory violates \eq{FL0} for a finite range of $\Delta$. 

 We have also checked cases with more than a single stack and more than one brane or antibrane per stack. These correspond to a higher-order polynomial $W'(x)$ \cite{Aganagic:2006ex}. Perhaps surprisingly, when one has exactly two stacks and $N>2$, and the branes are far apart enough that there is no tachyon, \eq{FL0} is satisfied. The light fields coming from the stabilized tachyon that became a problem before are now confined, as are the bifundamental states; only baryonic states are free, and they satisfy \eq{FL0}. However, this property is anecdotic, and it breaks down whenever one considers more than two stacks: already with three stacks one can violate \eq{FL0}, and with four or more stacks, it is possible to violate \eq{FL0} at large $N$.

The gauge kinetic function of the vectors can be computed exactly, see for instance \cite{Cachazo:2001jy,Aganagic:2006ex}. Of course, all the setups in this Section, except those with a single stack, have deformations (the parameter $\Delta$) that become dynamical one gravity is turned on.  
In this case, we can see explicitly why we should be wary of applying the reasoning behind FL in field theory situations with multiple $U(1)$'s. The multiple $U(1)$'s simply represent the fact that there is more than one brane stack. The stacks have opposite charges, and hence are attracted to each other; but since they wrap minimal size $S^2$'s, the systems is stable. The size of the $S^2$'s and the separations between the stacks are encoded in $W'(x)$, and correspond to non-normalizable modes in the supergravity description, acting as parameters \cite{Gukov:1999ya}. Once gravity is coupled, however, all these parameters become dynamical fields; the size of the $S^2$ will shrink, and the separation of the different stacks will decrease, since the moduli will not be stabilized. The vacuum we were looking at will be unstable, and the branes will coalesce and annihilate. We regard the multiple $U(1)$'s as a proxy that there are dangerous unstabilized moduli lurking around the corner, waiting to destroy the vacuum as soon as gravity is coupled again.

\section{Application to antibrane uplifting scenarios}\label{sec:app}

We have given stringy evidence for the FL bound by looking at non-compact models, which evade the problem of having to confront dS model building. We argued that in the QFT limit we can evade the FL bound by having decoupled sectors. But for compact models, ie at finite $M_p$, the FL bound should not be violated. This should imply that upon ``compactification", decoupled sectors couple through gravity and induce instabilities that lead to sufficient fast runaways, avoiding the application of the FL bound.

The two local models we considered were KPV \cite{Kachru:2002gs} and ABSV \cite{Aganagic:2006ex}. When both these mechanisms are used in compact models, KPV would correspond to anti-brane uplifting in some moduli-stabilised AdS vacuum a la KKLT \cite{Kachru:2002gs}, whereas ABSV would correspond to lifting via fluxes only, something that was suggested by Saltman and Silverstein \cite{Saltman:2004sn}. 
Since anti-brane uplifting is the most concrete scenario we analyse below to what extend it obeys the electromagnetic Swampland bounds
\footnote{The $U(1)$ gauge field we rely on for checking Swampland bounds comes from the anti-D3 brane as in our discussion of the flat space limit in the previous section. Note that this implies that the $U(1)$ is not further projected by orientifolds which can happen in fine-tuned situations \cite{Kallosh:2014wsa}. }. We will see that all bounds are satisfied in warped models (within the regimes of control) in the simplest settings. We then argue that a few extra complications, such as extra branes in the bulk, can violate the FL bound if one makes the usual assumption that the SUSY-breaking dark sector can be decoupled from the rest.

But before that, note there is a further (tower of) $U(1)$-vector(s) living down the throat. It comes from integrating $C_4$ along the $S^3$ (taking into account that parts of that tower is projected out by the orientifolds). These vectors are massive, but become massless in the limit of infinite throat size. Since those vectors turn out to be heavier than the Hubble scale the FL bound cannot be applied.

\subsection{Throats in compact models} 

To compute masses and charges of the lightest states charged under the anti-D3's U(1) gauge field we need some of the details of warped throat metric in compact models.

We will assume a KS-like throat in a compact CY as in picture \ref{fig: throat1} below.
\begin{figure}[ht]
	\centering
	\includegraphics[width=0.6\textwidth]{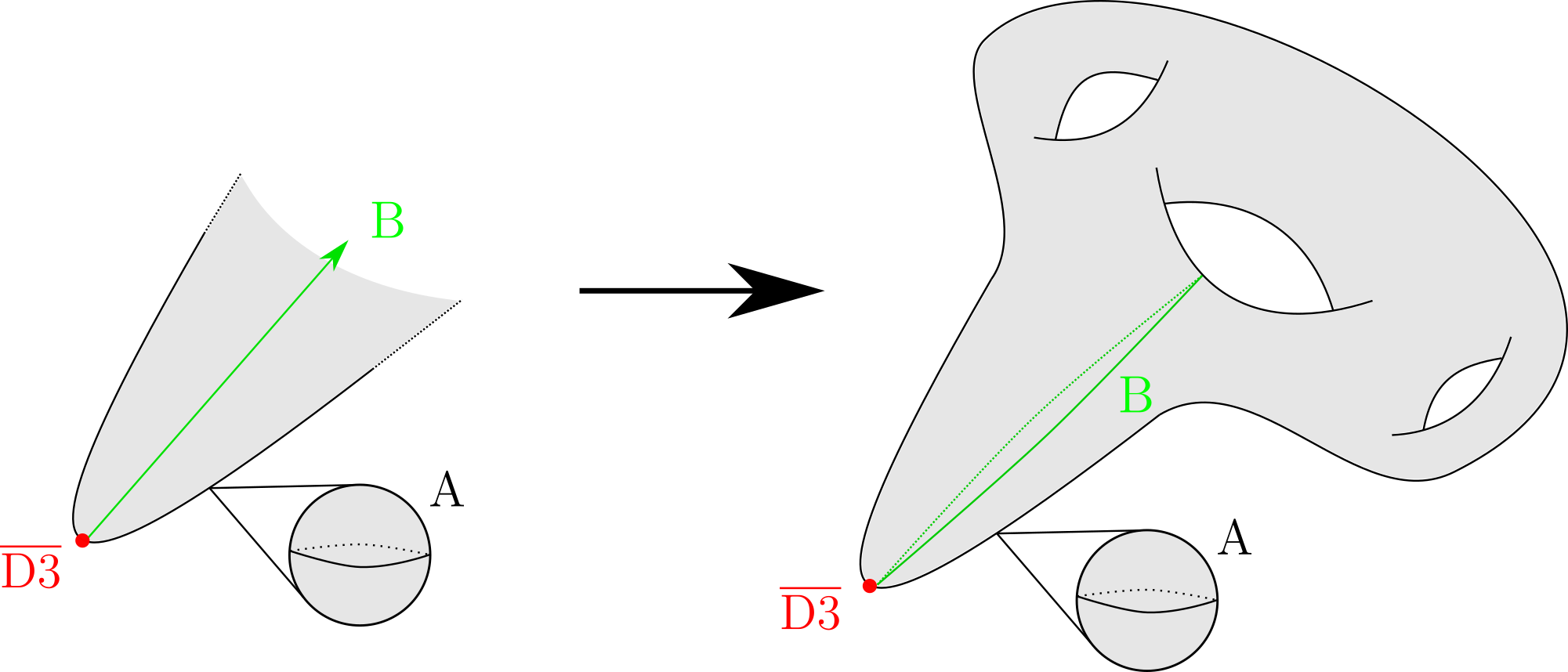}
	\caption{The KS throat and its compactification.}
	\label{fig: throat1}
\end{figure}
The 10D metric, in string frame, can be written as in the non-compact model (\ref{warpedCY}). At the tip the warpfactor reaches an exponentially small number:
\begin{equation}
e^{2A}\approx r_0^{-2} S^{2/3}\,, \quad r_0=\sqrt{g_s M}\,l_s\,.
\end{equation}
Here $S$ is the local conifold modulus and is stabilised at
$ S \sim \sqrt{\text{Vol}}\,\, e^{-2\pi\frac{K}{g_sM}} $,
where $K$ is the NSNS flux quantum piercing the throat B-cycle and $M$ the RR flux quantum piercing the A-cycle. At the same time the CY metric near the tip scales as 
$ds^2_6 \approx S^{2/3}$.
Hence the radius of the tip equals  $r_0$ and is fixed and independent of the volume. The total 10D metric near the tip becomes: 
\begin{equation}\label{tipmetric}
ds_{10}^2= e^{2A_0} \sigma^{1/2} g_4+ r_0^2\left[\frac12d\tau^2+d\Omega_3^2+\frac{\tau^2}{4} d\Omega_2^2\right],\qquad e^{2A_0}=r_0^{-2}e^{-4\pi\frac{K}{3g_sM}}\,. \end{equation}
The total volume is denoted as $\sigma=$Vol$^{2/3}$ and in case of a single K\"ahler CY $\sigma$ is the K\"ahler modulus. We will also need the physical volume $\mathcal{V}_T$ of the throat. In \cite{DeWolfe:2004qx} this was approximated to be
\begin{equation}
\int_T e^{-4A}\sqrt{\tilde{g}_6} = \mathcal{V}_T \sim (g_sMK)^{3/2}l_s^6\,.
\end{equation}	
In an approximation where most of the volume is coming from away from the throat, the Planck mass is \cite{Parameswaran:2020ukp}
\begin{equation}2\kappa_4^2=\frac{16\pi}{M_P^2}=\frac{g^2_s l_s^2}{2\pi \sigma^{3/2}}.\end{equation}
Recent investigations \cite{Gao:2020xqh, Carta:2021lqg} show that at least for the case of KKLT moduli stabilisation this condition is not attainable and most of the volume is in fact occupied by the throat. We have been careful in making sure that this does not effect the eventual bounds we have verified in this paper.

In Planck units\footnote{By which we mean  $ L \supset M_p^4 V$.} , antibranes contribute
\begin{equation}V_{\overline{D3}}\sim \frac{g_s^3}{4\pi} \frac{e^{4A_0}}{\sigma_0^2}, \end{equation}
to the 4d cosmological constant. Antibrane uplifting scenarios assume that $V_{\overline{D3}}$ is comparable to that of the negative AdS energy such that eventually the total on-shell potential is assumed to be
\begin{equation}
V = \alpha V_{\overline{D3}}\,,\quad \rightarrow \quad HM_p = \sqrt{\frac{\alpha \sigma_0}{g_s}}e^{2A_0}\,,
\end{equation}
where $0<\alpha<1$, but we do not assume that a finetuning $\alpha\rightarrow 0$ is possible and hence we take $\alpha$ to be order 1.  

To compact throat models we can associate a global and a local (ie warped down) KK scale. In string units they are $m^2_{KK} \approx \sigma^{-1/2}$ and $m^2_{\text{warped} KK} \approx e^{2A_0}\sigma^{-1/2}$. In total we have the following four distinct scales in four-dimensional Planck units,
\begin{equation} H\sim V^{1/2}=\frac{g_s^{3/2}}{\sigma_0} e^{2A_0} ,\quad m_{KK}^{\text{Bulk}}=\frac{\sqrt{\sigma_0}}{g_s},\quad m_{KK}^{\text{Throat}}=\frac{\sqrt{\sigma_0}}{g_s}e^{2A_0},\quad V^{1/4}\sim\frac{g_s^{3/4}}{\sqrt{\sigma_0}}e^{A_0} \end{equation}
The following hierarchy is then typically assumed:
\begin{equation} H< m_{KK}^{\text{Throat}}<V^{1/4}<m_{KK}^{\text{Bulk}},\end{equation}
which implies that the throat KK modes are very light and there is no real decoupling limit. Interestingly it is argued that the light throat modes only renormalise couplings and can essentially be ignored in the anti-D3 EFT \cite{Blumenhagen:2019qcg}.

\subsection{Swampland bounds for simple uplifts}
We will consider the same states as discussed in the previous section, namely string stretching from the anti-D3 towards the bulk (ending on D7 branes) and baryon particles from wrapped D3 branes. 

We also consider their magnetic counterparts being D1 branes that are attached to the anti-D3 brane and leave the throat, and D1 branes coming from a D3 brane wrapping the B-cycle. The first carries a unit of magnetic charge, the second $K$ units, with K the NSNS flux quantum in the B cycle. 

Let us start with the strings leaving the throat. The mass of the lightest strings can be estimated from their probe action:
\begin{equation}m_q\approx \frac{g_s}{\sigma^{1/2}} L_t M_p,\end{equation}
where $L_t$ is the length of a radial geodesic leaving the throat in a CY metric for the compact space with unit volume.

Perhaps confusingly $L_t$ is smaller for models with larger warping (ie big $K$). The reason is that the CY metric $ds^2_6$ (so with conformal factor taken out) scales as $e^{2A_0}$. We therefore expect that $L_t \sim e^{A_0}$, so it is exponentially small and the electric WGC is well obeyed.

Concerning the FL bound, we find that the scalings with $\sigma$ and the exponential warping drop out and the bound becomes:
\begin{equation}
2\pi L^2_t >> \sqrt{
	4\pi\alpha} e^{2A_0}\,. \label{ewe33}
\end{equation}
This can easily be satisfied if $L_t$ has a piece that does not scale as $e^{A_0}$. This can happen if the D7 branes are far enough away from the throat. Given recent developments this might not be possible \cite{Gao:2020xqh, Carta:2021lqg} and we need to be more careful in our estimates. Using the form of the 10D metric one expects
$L_t \sim  e^{-2\pi \frac{K}{3Mg_s}}$. This factor resides in the RHS of (\ref{ewe33}) as well, but there is an extra $r_0^{-2}$ suppression. Ignoring numerical factors this boils down to (string units)
\begin{equation}\label{FLstring}
(r_0)^{2} >> \sqrt{\alpha}\,,
\end{equation}
which is well satisfied if $r_0$ is large enough. The latter condition is a well-known requirement for stability of anti-brane uplifting \cite{Bena:2018fqc}.  Note how the FL bound becomes an expression independent of the total volume and hence the compactification effects.

Since a vibrating string generates a tower of modes, we need to verify the FL tower version (\ref{towerFL}) by taking into account the vibrational modes. Assuming the string has length $L_t$ in string units, we would like to consider how many vibrational states it has before reaching an energy scale $E_{max}$ at which the string will oscillate wildly, potentially leaving the throat region where this calculation might break down. Let us call this energy scale $E_{max} = N_{max}m_q$ where $m_q$ is the energy of the string at rest. We can estimate the number of states using Cardy's formula, $\rho(E)\approx l_s e^{\sqrt{c e^{-A_0} l_s \Delta E}}$, where we took into account that the local string scale is not $l_s^{-1}$ but $e^{A_0}\sigma^{1/4}l_s^{-1}$. From \eqref{towerFL}, we find for the pair production rate
\begin{equation}
\Gamma \sim e^{-A_0}\int^{E= N_{max} m_q}_{E=m_q} dE e^{\sqrt{c e^{-A_0}\sigma^{-1/4}} l_s E} e^{-E^2 / q g H M_p }\,.
\end{equation}
We wish to impose that this pair production rate is supressed. To be safe we will even overestimate the integral, replacing $E^2$ with its rest mass energy $m=m_q$. Really, the heaviest state we consider has a mass $N_{max} m_q$. This changes the results by an $\mathcal{O}(N_{max})$ numerical factor. We have
\begin{equation}
\label{eqtowerboundwprefactor}
\Gamma <\sim  e^{- m_q^2 / q g H M_p } \left[\frac{2 e^{\sqrt{c e^{-A_0}\sigma^{-1/4}l_s N_{max} m_q}}(\sqrt{c e^{-A_0}\sigma^{-1/4}l_s m_q}-1)}{ce}+\frac{2}{c }\right]\,,
\end{equation}
where we further dropped a negative term that is subleading, again overestimating the pair production rate. 
To have the  rate supressed, we must demand \eqref{eqtowerboundwprefactor} $\ll 1$ and focussing only on the exponential factors in this bound we get:
\begin{equation}\label{FLstring2}
(r_0)^{2} - \sqrt{cN_{max}\alpha r_0} >> \sqrt{\alpha}\,,
\end{equation}
A natural guess for $N_{max}$ would be that it roughly equals $2$ since once the vibrational energy becomes of order the rest energy the string wildly oscillates and potentially leaves the throat, invalidating the assumptions. Hence we do not expect to violate \eqref{FL0}.

Next we consider the wrapped D3 states. A $D3$-brane wrapped on the $S^3$ has charge $M$ due to a Freed-Witten anomaly: $M$ strings leave the D3 and attach to the anti-D3. From the 10d metric (\ref{tipmetric}) it follows that the 4d mass of such $D3$ brane in string units  is  
\begin{equation} m_{D3}\approx g_s^{-1} \sigma^{1/4}e^{A_0} r_0^3.\end{equation}
The flat space WGC inequality then becomes
\begin{equation}
e^{-\frac{2\pi K}{3g_sM}}<\sqrt{\frac{4\pi\sigma}{g_s}}\,.
\end{equation}
This is trivially satisfied.\footnote{Note that flat space WGC is applicable if $m^4>>H^2M_p^2$. This is the case.} The FL bound (with $q=M$) gives
\begin{equation}
r_0^4 >\sqrt{4\pi\alpha}g_s\,.
\end{equation}
Both the validity of the supergravity description and stability \cite{Bena:2018fqc} require $r_0>>1$, so the FL inequality is satisfied as well and again is volume independent.

Just like the string we discussed before this state is part of a tower of charged states that lives down the throat. Hence, local excitations, which manifest themselves as different particles in 4D, will also be light and we need to apply the tower version of the FL bound (\ref{towerFL}). We investigate this now for states generated by the wrapped D3's moving along its homology class; an effective motion inside the local B-cycle (transverse to the wrapped A cycle). We focus on the 7d dynamics of its center of mass as this will be a good approximation at the bottom of the throat, as long as the eigenvalues of the energy excitations are smaller than the mass $m$ of the brane. This mass is obtained from the kinetic term,
\begin{equation} T_{D3} \int_{D3} dV,\quad dV_3= \omega_3 \wedge ds_7= \sqrt{g_{\mu\nu} \dot{x}^\mu \dot{x}^\nu} \omega_3\wedge dt,\end{equation}
where $ds_7$ is the seven-dimensional obtained from restricting the 10d metric to the product of B-cycle and 4d transverse space.  The only force comes from the fact that the $D3$ emits $M$ F1's. We assume these F1's are straight and in their fundamental state, and that they contribute to the energy an amount comparable to their tension times their length.

The resulting potential term (in string units) is
\begin{equation} M \int ds_2= M e^{A} \ell(r) dt,\end{equation}
where $\ell(r)$ is the geodesic length from the bottom of the throat to a position at radius $r$, measured in the physical metric.

One eventually finds that fluctuations near the tip are described by a non-relativistic linear potential problem in quantum mechanics. Consequently, the energies are
\begin{equation}
E_n = m \left(1 + \left(n\,\frac{g_s}{r_0^4}\right)^{2/3} \right)\,.
\end{equation}
At $n_* = \frac{r_0^4}{g_s}$ the approximation certainly breaks down since the kinetic energy becomes order mass and non relativistic quantum mechanics does not apply.  The tower FL bound demands that $\Gamma << 1$. Since all masses up to $n_{*}$ are of the same order we roughly get (going all the way to $n_*$):
\begin{equation}
\frac{r_0^4}{g_s}e^{-\frac{r_0^4}{g_s}}< 1\,.
\end{equation}
Clearly the exponent wins and the inequality is satisfied. 

The magnetic states can be treated along the identical lines and we have verified that they all satisfy the WGC and FL inequalities upon assuming the usual conditions of long throats and weak coupling. 

One can wonder about the deeper lying reason the FL bound is satisfied in the above context. In section \ref{sec:rev} we derived that 4D particles made from small spherical membranes in the theory tend to be light charged states with a mass scale set by the vacuum energy. Then we derived a lower mass bound of a kind similar to the FL bound but with a  different (weaker) dependence on the gauge coupling. We now explain that the D3 particle can really be thought of as a membrane-particle in the above context which we believe forms the underlying reason the FL bound is satisfied. As explained by KPV \cite{Kachru:2002gs} the spacetime membrane that mediates the decay is an NS5 brane wrapping the whole A-cycle and its tension is therefore
\be
T \sim \text{Vol}(S^3) g_s^{-2}
\ee
in warped down string units.  The baryon particle on the other hand has a mass; $
m \sim g_s^{-1} \text{Vol}(S_3)$, in the same units. So up to a $g_s$ dependence this is like the relation between particle mass and membrane tension for membrane-particles, which is enough to find a lower bound on the mass as a consequence of vacuum stability. To make the connection more precise we have to argue that the baryon puffs into a small spherical membrane state in 4D. In fact this is easily done once one realises the puffing is almost identical to the polarisation of the anti-D3 into a spherical NS5 wrapping a contractible $S^2$ on the A-cycle \cite{Kachru:2002gs}. The main difference is that now the polarisation occurs in 4D non-compact space as opposed to a process inside the compact dimensions. The coupling $F_2\wedge C_4$ with $F_2$ some worldvolume flux on the NS5 membrane guarantees it induces the same D3 charge on the A-cycle. Similary the usual DBI worldvolume coupling takes care of the tension. 

\subsection{Swampland conflicts with warped uplifts}

We found that the longer the throat, the larger the CY volume and the weaker the coupling, the better all bounds (electric $\&$ magnetic WGC and FL) are satisfied,  and this makes sense given the decoupling limit we analysed earlier. That these Swampland constraints are automatically satisfied merely reflects the fact that the throat is very much decoupled from the bulk CY and, in the case of FL, it also relies on the fact that the throat model is KPV which, as explained in the previous section, satisfies a noncompact version of FL.

Since satisfying these Swampland bounds only speaks to the consistency of the throat, it does not constitute evidence for the consistency of the uplift procedure in those limits. One reason is that we do not expect these limits to be parametrically attainable in any globally consistent scenario, as hinted in a number of works \cite{Ooguri:2018wrx, Junghans:2018gdb, Banlaki:2018ayh}. But more to the point here, FL is really a global bound, and even though it is satisfied in the throat, we will see we can violate it by adding decoupled sectors far away from the throat that are usually argued to be harmless. In the non-compact setups discussed earlier, we argued that decoupled sectors typically lead to runaways, and  the non-compact version of FL should not apply; but the black hole arguments behind FL suggests it must apply to \emph{any} compact model, and so a violation signals a problem with the compactification.

Let us be more concrete. The premise of anti-brane uplifting down a warped throat is that the anti-brane SUSY breaking decouples sufficiently from the bulk CY. In the bulk CY one can then attempt to engineer interesting particle physics properties. We now argue that such decoupling cannot be there as it can violate the FL bound. 

\begin{figure}[ht]
	\centering
\includegraphics[width=0.4\textwidth]{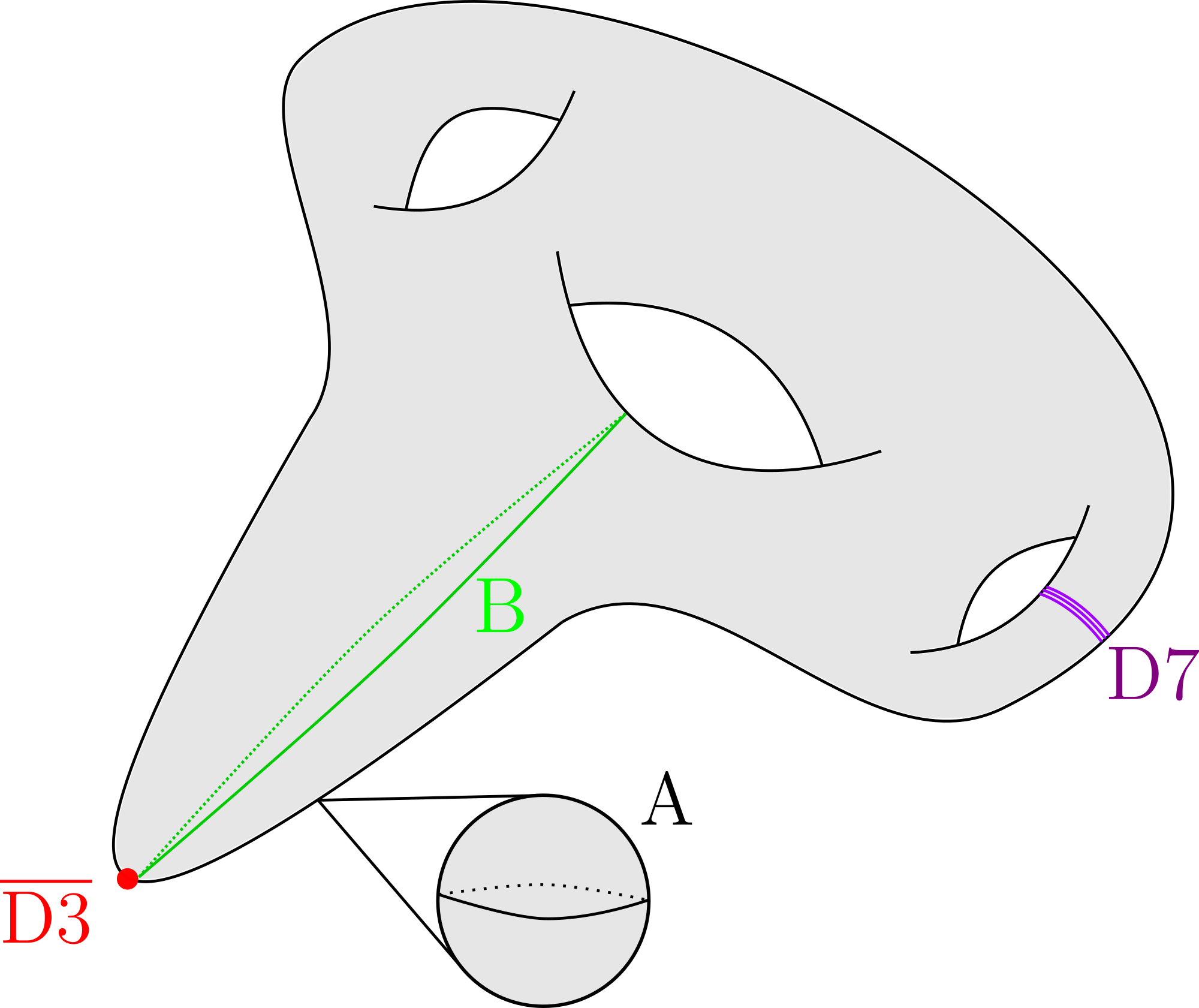}
\caption{In purple we have depicted D7 branes wrapping a distant 4 cycle in the bulk. Such 7-branes can support chiral matter and in the usual assumption that the SUSY breaking down the throat decouples from the bulk, one finds a conflict with the FL bound.}
\label{fig:chiral}
\end{figure}

To do this, one may consider for instance local F-theory constructions that generate chiral matter \cite{Donagi:2008ca,Beasley:2008dc} .  In our setup one can think of 7-branes wrapped on a distant divisor as in picture \ref{fig:chiral}. This divisor is not threaded by flux, so it contributes a 4d $\mathcal{N}=1$ sector to the low-energy effective field theory, involving massless charged fields. To be fully concrete, take the toy model of \cite{Beasley:2008dc} where a stack of $E_6$ 7-branes wraps a Hirzebruch surface $F_1$. As analyzed there, at low energies the system is described by a 4d $\mathcal{N}=1$ theory with gauge group $\text{Spin}(10)\times U(1)$.  The matter fields in the theory are the $\mathcal{N}=1$ vector multiplet in the adjoint, as well as a chiral spectrum of massless fields in the $\mathbf{16}_{-3}$ and $\overline{\mathbf{16}}_{+3}$ of $\text{Spin}(10)$, with net chirality
\begin{equation}n_{\mathbf{16}}-n_{\overline{\mathbf{16}}}=3(2a+b),\end{equation}
where $a,b$ are integers of opposite signs quantifying the gauge bundle on the 7-brane stack.  In this theory, the fermions in the $\mathbf{16}_{-3}$ and $\overline{\mathbf{16}}_{+3}$ violate FL for the $U(1)$ factor. On the other hand, it is often (but not always) the case that a $U(1)$ factor like this gets a mass from the Green-Schwarz mechanism \cite{Ibanez:2012zz}; in this case, the $U(1)$ is rendered massive, and FL does not apply. FL requires that this always happens for any local model similar to this one, for a global model with positive vacuum energy.

On the other hand,  the $\text{Spin}(10)$ gauge fields can also violate FL if there are light gluons, i.e. if the gauge fields do not confine in the deep IR. Using the NSVZ formula \cite{Novikov:1983uc} (with vanishing anomalous dimensions, which is justified if the volume of the $F_1$ is large so that the 4d theory is weakly coupled), and assuming that nonchiral matter will pick a mass,  the theory will run towards weak coupling in the IR if 
\begin{equation} 3T_{\text{Adj}} - \left(T_{\mathbf{16}} \vert n_{\mathbf{16}}-n_{\overline{\mathbf{16}}}\vert \right)=6\, (4-\vert2a+b\vert)<0,\end{equation}
where $T_{\mathbf{R}}$ is the Dynkin index in representation $\mathbf{R}$ (which are 8 and 2 for the adjoint and the $\mathbf{16}$, respectively).  For e.g. $a=-b=3$, the $\text{Spin}(10)$ theory is weakly coupled in the IR and deconfined, thus providing a sector that violates FL. The violation is twofold: on one hand, the existence of massless $\text{Spin}(10)$ gluons directly contradicts FL.

This picture ignores the SUSY-breaking backreaction sourced by the antibrane at the end of the throat. A priori, this could trigger e.g. Higgsing, breaking $\text{Spin}(10)$ completely and avoiding a contradiction with FL, if the induced masses for the gauge bosons are above the Hubble scale. But this is precisely the point we want to emphasize: the decoupling scenario, where the antibrane can be safely neglected, is in contradiction with Festina Lente. FL is a global constraint; satisfying it means turning on interactions between far away and seemingly decoupled sectors. 
Perhaps this is in the same spirit as the recent findings of \cite{Gao:2020xqh} and \cite{Carta:2021lqg} where the limit of control on anti-brane SUSY breaking in a KKLT scenario implies either that the bulk becomes completely singular, or if resolved, it affects the whole bulk geometry. 

At the risk of belaboring the point, we emphasize again that the scenario we have depicted here, where we imagine one can find a distant divisor where additional $D7$ branes can be wrapped safely, is far from concrete. Indeed, we do not believe it can happen! But it is a plausible scenario, similar to the KKLT or LVS scenarios \cite{Kachru:2003aw,Balasubramanian:2005zx}, where we see that by combining several locally consistent ingredients in a seemingly reasonable way we end up with an effective field theory which violates Swampland principles. The fact that by following the same logic as a generic antibrane uplift model we find inconsistencies hints at the possibility that the basic logic itself is flawed due to the lack of decoupling whenever gravity is involved.

\section{Conclusions}\label{conclus}

 In this paper we have extended the Festina Lente bound \cite{Montero:2019ekk} for charged particles in quasi-dS space in various directions, including its formulation in higher dimensions, a magnetic version, an extension with multiple gauge and scalar fields, even including runaway potentials. This allowed us to explore some of its implications for particle physics and string phenomenology. We furthermore provided some evidence for it coming from its nontrivial decompactification limit. From this we drew several interesting conclusions, which we summarize below. 

The formulation of the magnetic WGC in dS space leads to a lower (non-zero) bound on the gauge coupling in terms of the ratio between Hubble and Planck scale \cite{Huang:2006hc} and we found that
\begin{itemize}
    \item The same condition, $g>H/M_p$, can be found from applying the FL bound to the WGC particles.
\end{itemize}

If we combine the FL proposal to dimensional reduction, we have shown, in the case of a circle compactification, that 
\begin{itemize}\item A de Sitter vacuum must be minimally scale-separated, $M_{\text{KK}}\gtrsim \Lambda^{1/2}$.\end{itemize}
What we mean precisely is that if the KK vacuum (stabilized due to contributions of Casimir energies) of a $(d+1)$-dimensional theory with positive vacuum energy itself has a positive vacuum energy, then it must be scale-separated in the sense above.  This behavior is opposite to that of known string constructions with negative cosmological constant. 

This conclusion came out of applying FL to the KK photon. We also analyzed how the FL inequality behaves under dimensional reduction, i.e. when a $(d+1)$-dimensional theory with positive vacuum energy an $U(1)$ gauge fields is compactified on a circle. We found that, if the radius of the $S^1$ is left unstabilized,
\begin{itemize}\item FL is preserved under dimensional reduction.\end{itemize}
That is, the lower-dimensional theory obeys FL if and only if the higher-dimensional one does as well. In the case where the size of the circle is stabilized by Casimir energy, FL is easily, but not automatically, satisfied. It is possible that analyzing this interplay in detail could result in bounds on the spectrum of light fields of the $(d+1)$-dimensional theory, as has been done with other Swampland constraints \cite{Heidenreich:2015nta}. 

 The main appeal of an inequality like FL is that it has the potential to produce interesting phenomenological constraints. FL requires that when we have a quasi-dS background
\begin{itemize}
\item All non-abelian gauge fields are confined or broken spontaneously at a scale above Hubble, just like the real world $SU(3)$ and $SU(2)$ fields.  
\item We have found that FL can ameliorate the hierarchy problem, has potential connections to neutrino physics, and can be used to exclude a symmetric minimum at $\Phi=0$ in the Higgs potential. This is an experimental prediction, even if a modest one. 
\item In a particular scenario, the FL proposal also leads to neutral particles with a relation $m^4\sim V$. It is natural to speculate that these may be related to neutrinos, although the current picture is qualitative at best.
\end{itemize}
We have only started to scratch the surface in the arena of phenomenological implications of the FL proposal, which can lead to nontrivial restrictions for dark matter or inflationary scenarios.

Concerning dS model building in supergravity we observed that 
\begin{itemize}\item Bottom-up gauged supergravity dS constructions tend to violate FL.\end{itemize}
This extends some recent observations made in \cite{Cribiori:2020use, Cribiori:2020wch}.
Although there was no reason to expect that these bottom-up constructions could not be embedded in string theory, FL puts a sharp obstacle to this being the case and we found a link with the no-global symmetry conjecture.

We have finally explored potential evidence for FL in string theory, coming from the fact that it has a nontrivial flat space limit, where gravity is decoupled. There, FL becomes a statement about the spectrum of charges and particles of metastable nonsupersymmetric vacua of  quantum field theory arising from brane constructions in string theory. One needs to make several caveats in the application of FL to systems where gravity is decoupled, in order to ensure stability under small coupling to gravity. But once these are made, we find that the resulting metastable vacua do comply with what we propose might be a flat-space version of the FL bound. In particular, and most importantly for phenomenology,
\begin{itemize}\item The $U(1)$ worldvolume gauge field of a $\overline{D3}$ brane at the tip of a Klebanov-Strassler throat satisfies the non-compact version of FL.\end{itemize}
We have used this result to check that
\begin{itemize}\item Both FL and the WGC are satisfied by the  $\overline{D3}$ $U(1)$ in antibrane uplift scenarios. \end{itemize}
This is essentially because in the antibrane uplift scenario, the antibrane lives in a decoupled throat, and so the problem reduces to the noncompact one, so that Swampland constraints are satisfied automatically. Even if the antibrane is OK with Festina Lente, the decoupling of the throat and the rest of the compactification space is problematic. We gave explicit examples of local sectors outside of the warped throat which, if present, would violate the FL inequality. From the low-energy point of view, there seems to be nothing wrong with these decoupled sectors, and indeed, they are often included to introduce additional gauge fields or chiral matter; but if FL is correct, the possible existence of these sectors would raise questions for the ability to incorporate uplift scenarios in a compact setup. 

As we have seen, if correct, the FL proposal leads to very rich implications both in particle physics and string phenomenology. We have explored some of the more salient ones, but a more systematic study is desirable. It is imperative to check whether FL is satisfied in some recent new proposals for dS model building in string theory, such as \cite{DeLuca:2021pej, Banerjee:2018qey}. This can be used to check both the FL statement, the principles behind it, and the dS constructions themselves.

We can also do better with the conjecture itself. It would be nice to check the prediction of $\sqrt{6}$ for the $\mathcal{O}(1)$ coefficient for the FL bound, which we obtained by demanding agreement with the analysis of magnetic Nariai black holes, against a direct calculation of black hole decay in the strongly coupled Schwinger regime. But there are plenty other questions to address.  How is FL modified in the presence of other non-trivial interactions? Can the conjecture be related to general properties of entanglement entropy in quasi-dS space \cite{Aalsma:2020aib,Aalsma:2021bit,Aalsma:2021kle} and black holes, just like the WGC can \cite{Montero:2018fns}? Is it possible to come up with complementary evidence in string theory that can support or disprove the conjecture? What are the minimal requirements that FL puts on inflation? Can the argument be generalized to other kinds of black holes, possibly to directly produce constraints on the scalar potential? These are all interesting questions we hope to revisit in upcoming work.

At this point it is useful to summarize how much evidence we have for the FL bound. First, the original argument \cite{Montero:2019ekk} demands that black hole evolution behaves well in the sense that black holes should be able to evaporate completely. This is how the ordinary electric WGC was found in flat space and in dS space it simply leads to a second inequality. Second, we can equally reach the same conclusion by applying the magnetic WGC and demanding the cut-off is above the Hubble scale as we have explained. This is evidence from an a-priori unrelated argument. This shows the usual behavior of Swampland conjectures; they form a tied self-consistent web. Thirdly we have found circumstational evidence from string theory and supergravity models. For instance violating the no-global symmetry conjecture in supergravity allows constant FI terms \cite{Komargodski:2009pc}, which directly tend to violate the FL bound. Similarly local stability constraints of anti-brane uplifting point exactly in the direction of the FL bound, whereas the assumption of decoupling SUSY-breaking from the bulk violates it. This makes much sense in light of general Swampland ideas concerning a lack of decoupling due to light towers. So it supports the picture in two directions; note that decoupling is not a necessary requirement for consistently achieving dS vacua, but it does make it much easier to verify its consistency. So we argue that if consistent anti-brane uplifting is possible, the model is highly constrained. Lastly, we have argued that even pure QFT models obtained from a decoupling limit potentially support this picture. 

Finally, FL and all of its far-reaching consequences are the result of applying ``black hole arguments'' beyond the controlled realm of supersymmetric string compactifications. This illustrates why is it so important to understand these black hole arguments fully; They not only provide nice insight or explanations as to why Swampland conjectures are true, but also they tell us how to proceed in situations where string theory cannot directly be of help. It is an exciting prospect to consider what other constraints on low-energy EFTs might come from bold applications of similar well-established general principles!

\subsection*{Acknowledgements} We thank Arthur Hebecker, Fotis Farakos, Julian Mu\~{n}oz, Matt Reece, Irene Valenzuela, Luis Iba\~{n}ez, Eduardo Gonzalo, Angel Uranga, Matt Strassler, Gary Shiu, Pablo Soler and Susha Parameswaran for valuable discussions and comments on the manuscript. TVR and MM also thank Juan Joya Borja for moral support. This work is supported by the KU Leuven C1 grant ZKD1118C16/16/005 and the FWO fellowship for sabbatical research. The work of CV and MM was supported by te National Science Foundation under Grant No. NSF PHY2013858, as well as by a grant from the Simons Foundation (602883, CV).

\appendix 

\section{Multi-field FL bound}\label{App:multi}
String compactifications typically generate multiple $U(1)$ gauge fields and so we are naturally lead to considering Nariai black holes charged under the multiple gauge groups. So we seek here the extension of the FL inequality to that case.  Our guiding principle will be multi-field covariance. 

Consider the general Lagrangian
\begin{equation}
e^{-1}\mathcal{L} = \frac{M^2}{2}\mathcal{R} -\tfrac{1}{2}G_{ij}(\phi)\partial\phi^i\partial\phi^j -\tfrac{1}{4} f_{AB}(\phi)F^AF^B - V(\phi)\,. 
\end{equation}
where we ignore theta angles and assume that the scalars are not charged for simplicity. The FL bound (\ref{FL0}) then generalises to
\begin{equation}
m^4 >> q_Aq_B (f^{-1})^{AB} (M_pH)^2 \,.
\end{equation}
Where $q^A$ is the charge vector for a particle with mass $m$. For magnetic or even dyonic particles the further extension is then also easily guessed:
\begin{equation}
m^4 >> \left((f^{-1})^{AB}q_Aq_B  + f_{AB}p^Ap^B \right)(M_pH)^2 \,,
\end{equation}
with $q_A,p^B$ the electric-magnetic charge vector.

In case the scalar potential has only a runaway then there can still be charged Nariai solutions and the FL bound can be applied under conditions that generalise the single field condition (\ref{dScond0}). Going through the identical analysis done in \cite{Montero:2020rpl} one finds the following consistency condition 
\begin{equation}
G^{ij}\frac{\partial_i V \partial_j V}{V^2} \leq G^{ij}\frac{\partial_i \mathcal{Q}^2 \partial_j \mathcal{Q}^2}{(\mathcal{Q}^2)^2}\,,
\end{equation}
where we defined:
\begin{equation}
\mathcal{Q}^2 = (f^{-1})^{AB}Q^e_AQ^e_B + f_{AB}Q_m^AQ_m^B\,.
\end{equation}
The $Q^e_A, Q^B_m$ are the black hole electric and magnetic charges. After some algebra this can be shown to become $Q$-independent and one obtains the following generalisation of (\ref{dScond0}):
\begin{equation}
G^{ij}\frac{\partial_i V \partial_j V}{V^2} \leq  - G^{ij} \partial_i f_{AB} \partial_j f^{AB}\,.
\end{equation}
The scalar positions in the above inequality are determined by the following equations \cite{Montero:2020rpl}:
\begin{align}
&\partial_i V + \frac{1}{2 R^4}\partial_i \mathcal{Q}^2 = 0\,,\\
& R^4 = - \frac{G^{ij}\partial_iV\partial_j\mathcal{Q}^2}{2G^{ij}\partial_i V\partial_j V}\,.
\end{align}
With $R$ the radius of the Nariai solution. A final consistency condition for the existence of the solution is that $V\mathcal{Q}^2<2$.

\section{Stability of \texorpdfstring{$d$}{d}-dimensional Nariai solutions with runaway potential}\label{App:higherd}
In this Appendix, we derive \eqref{ineqd} of the main text, generalizing the analysis in \cite{Montero:2020rpl}. We start with a $d$-dimensional Einstein-Hilbert-scalar system, with action (we work in Planck units $8\pi G_d=1$)
 \be S= \int\sqrt{|g|}\Bigl(\frac{1}{2} \mathcal{R}-\tfrac{1}{2}(\partial\phi)^2 -\tfrac{1}{4}f(\phi) F_{\mu\nu}F^{\mu\nu} - V(\phi)\Bigr),\label{act0}
\ee
which is a consistent truncation of the full theory. We look for $dS_2\times S^{d-2}$ solutions, supported by an electric field\footnote{Magnetic solutions produce higher-dimensional de Sitter configurations which are perturbatively unstable, as analyzed in \cite{Montero:2020rpl}. Only for $d=4$ magnetic solutions provide a viable Nariai branch, too.}
\begin{equation} \int_{S^{d-2}}(f *F)=\Omega_{d-2}Q_e.\end{equation}
where $\Omega_{d-2}$ is the volume of unit radius $S^{d-2}$. Following \cite{Montero:2020rpl}, we choose a metric
\begin{equation} ds^2= h(t) [-dt^2+dr^2]+ R^2d\Omega_{d-2},\end{equation}
and electric field
\begin{equation} F=E\, dr\wedge dt,\quad E= Q_e\frac{h}{f R^{d-2}}=\frac{h \mathcal{Q}}{\sqrt{f}R^{d-2}}.\end{equation}
Plugging back this ansatz in the action \eq{act0}, we obtain the following effective action in one dimension (we neglect $r$-dependence)
\begin{equation}S=\int dt\left[(d-2)(d-3) R^{d-4}(h-\dot{R}^2)-R^{d-2}\frac{d}{dt}\left(\frac{\dot{h}}{h}\right)+h R^{d-2}\left(\frac{\dot\phi^2}{2}-V\right)-\frac{h \mathcal{Q}^2}{2R^{d-2}}\right]\end{equation}
For $d=4$, this reduces to the result of \cite{Montero:2020rpl} for electric fields only\footnote{It seems that the kinetic term for $R$ has the wrong sign, but the dynamics is more complicated since the equation of motion for $h$ provides a constraint.}. The equation of motion for $R^{d-2}$ is (taking into account that we are looking for static solutions, and thus set $\dot{R}=\dot\phi=0$)
\begin{equation}\mathcal{R}_2=\frac{1}{h}\frac{d}{dt}\left(\frac{\dot{h}}{h}\right)=V-\frac{\mathcal{Q}^2}{2R^{2d-4}}-\frac{(d-3)(d-4)}{R^2}.\label{eomR}\end{equation}
The equation of motion for $h$ is 
\begin{equation} V R^{d-2}+\frac{\mathcal{Q}^2}{2R^{d-2}}-(d-2)(d-3) R^{d-4}=0,\label{eomh}\end{equation}
and the equation of motion for $\phi$ is
\begin{equation}0=V'+\frac{(\mathcal{Q}^2)'}{2R^{2(d-2)}}.\label{eomphi}\end{equation}
Importantly, notice that for $d=3$, \eq{eomh} has no solution. For $d>3$, we can substitute \eq{eomphi} and \eq{eomh} into \eq{eomR}, to obtain
\begin{equation} \mathcal{R}_2= \frac{2V}{d-2}\left((d-3)+ \frac{V'}{V}\frac{\mathcal{Q}^2}{(\mathcal{Q}^2)'}\right).\label{ep0}\end{equation}
Thus, existence of an electric Nariai solution requires that $V'/V$ and $f'/f$ have opposite signs, and also that
\begin{equation} \left\vert \frac{V'}{V}\right\vert \leq  (d-3)\left\vert \frac{f'}{f}\right\vert.\end{equation}
This is consistent with the analysis in Appendix C of \cite{Montero:2020rpl}, by taking the dual $p=d-2$-form potential, although strictly speaking the derivation in \cite{Montero:2020rpl} does not work for the two-dimensional case. 

To study stability of the Nariai solution, we only need to consider the $\phi$ field -- the radial mode is fixed by the Hamiltonian constraint of GR \cite{Montero:2020rpl} --. We obtain that stability is equivalent to
\begin{equation} V''(\phi )\geq \frac{V'}{f'} \left( f''(\phi )-\frac{2 f'^2(\phi )}{f(\phi )}\right).\end{equation}
Taking $V = V_0 e^{\delta \phi}$ and $f = f_0 e^{\gamma \phi}$, One gets
\begin{equation}
    \delta^2 \geq - \frac{\delta}{\gamma} \gamma^2\,,
\end{equation}
which is automatically satisfied when both $\delta$ and $\gamma$ have the same sign.

\bibliographystyle{JHEP}
\bibliography{FLrefs}

\end{document}